\documentclass[%
 reprint,
 amsmath,amssymb,
 aps,
]{revtex4-1}

\usepackage{graphicx}
\usepackage{dcolumn}
\usepackage{float}
\usepackage{bm}

\begin{document}

\preprint{APS/123-QED}

\title{Understanding Financial Market States Using Artificial Double Auction Market}

\author{Kyubin Yim}
\email{kyubin48@gmail.com}
 \affiliation{Nonlinear and Complex System Laboratory,Department of Physics, POSTECH, Pohang 790-784, Republic Of Korea}
\author{Gabjin Oh}%
 \email{phecogjoh@gmail.com}
\affiliation{Division of business administration, Chosun Univesity, Gwangju 501-759, Republic Of Korea}%

\author{Seunghwan Kim}
\email{swan@postech.ac.kr}
\affiliation{Nonlinear and Complex System Laboratory,Department of Physics, POSTECH, Pohang 790-784, Republic Of Korea}
\date{\today}

\begin{abstract}
The ultimate value of theories of the fundamental mechanisms comprising the asset price in financial systems will be reflected in the capacity of such theories to understand these systems. Although the models that explain the various states of financial markets offer substantial evidences from the fields of finance, mathematics, and even physics to explain states observed in the real financial markets, previous theories that attempt to fully explain the complexities of financial markets have been inadequate. In this study, we propose an artificial double auction market as an agent-based model approach to study the origin of complex states in the financial markets, characterizing important parameters with an investment strategy that can cover the dynamics of the financial market. The investment strategy of chartist traders after market information arrives should reduce market stability originating in the price fluctuations of risky assets. However, fundamentalist traders strategically submit orders with a fundamental value and, thereby stabilize the market. We construct a continuous double auction market and find that the market is controlled by a fraction of chartists, $P_{c}$. We show that mimicking real financial markets state, which emerges in real financial systems, is given between approximately $P_{c}=0.40$ and $P_{c}=0.85$, but that mimicking the efficient market hypothesis state can be generated in a range of less than $P_{c}=0.40$. In particular, we observe that the mimicking market collapse state created in a value greater than $P_{c}=0.85$, in which a liquidity shortage occurs, and the phase transition behavior is $P_{c}=0.85$.
\end{abstract}

\pacs{Valid PACS appear here}
\maketitle


\section{Introduction}
The study of asset pricing has a long history in the financial markets based on its crucial role, and understanding its fundamental mechanism is related to the investor behavior. No theory in the fields of economics and finance can explain all aspects of the price mechanism because the financial markets, which are one of the most complex systems, are characterized by various market states such as normal and abnormal states. To fully understand the price dynamics of risky assets, we must know the nature of the diverse states of financial markets. Two celebrated scholars in economics, Eugene F. Fama and Robert J. Shiller, received the 2013 Nobel Prize in Economic Sciences for their work that furthered the understanding of fundamental features of asset pricing in the financial markets \cite{nobeleconomic2013}. Eugene F. Fama proposed the Efficient Market Hypothesis (EMH), which is widely accepted as the foundational economic theory of asset pricing models, including pricing of both underlying assets and options related to such assets \cite{fama,assetprices}. EMH assumes that all investors in the financial markets are rational and posits that these investors cannot predict future market prices using past financial market information. However, Robert J. Shiller proposed `behavioral finance', which includes irrational traders who can predict future market prices using past information in the price movement of risky assets or by means of the psychological effects of investors \cite{shiller}. Behavioral finance has been further developed by many economists and is well established in mainstream financial economics \cite{behavioral_finance1,behavioral_finance2,behavioral_finance3,behavioral_finance4,behavioral_finance5,behavioral_finance6,behavioral_finance7}. Employing both EMH and behavioral finance, many market microstructure models have been developed \cite{inventorymodel1,inventorymodel2,inventorymodel3,informmodel1,informmodel2,informmodel3,informmodel4,strategicmodel1,strategicmodel2}. In addition, studies at the micro level that utilize high-frequency data from the financial markets have contributed to developing asset pricing and market microstructure field \cite{mandel,gopi,plerou1,maslov,cont}.\\
Unfortunately, the essential assumption of EMH is totally different from that of behavioral finance. In other words, most research on asset pricing models can explain some of the features of various aspects of the market by making specific assumptions regarding major factors, such as investors and the dynamics of risky assets. Thus, the difficult of fully understanding the asset price mechanism is grounded in the fact that a variety of independent theories contribute to complete asset pricing models, and fully understanding asset pricing has been possible only by means of integrated models that map both theories. Recent advances in understanding the stylized facts of asset prices in financial markets using an agent-based model (ABM) have shed light on factors (1) that prompt us to wonder whether the heterogeneity of investors is more important than representative agents in this context and (2) how interactions among traders affect the fundamental features of asset prices. As an understanding of asset prices that considers that traders are heterogeneous, ABM has recently emerged as an alternative approach in asset pricing models. 
The properties of ABM can be distinguished from other methodologies as follows: (i) ABM provides the linkage from the micro investor level to the macro market level and (ii) ABM can generate artificial data that include different market scenarios and heterogeneous agent assumptions. Therefore, an artificial stock market has been created with various scenarios that include not only heterogeneous agent types from zero intelligence models to multi-agent models but also various trading mechanisms, from market-clearing systems to order-driven markets \cite{hommes,boswijk,hommes2,mg,santafe,lux,lux1,chen,spinmodel,chiarella1,bartolozzi,miller,chiarella2,bouchaud,yamamoto}.\\
To illustrate an alternative asset pricing model reflecting the various market states of real financial markets, we develop an artificial double auction market (ADAM) constructed with heterogeneous agents, including fundamentalists and chartists, with a  double auction market as its trading system.\\
The remainder of this paper is organized as follows. In the Model section, we describe the ADAM. The results section describes and discusses the results of the data generated by the model. The Discussion and Conclusions summarizes the paper and proposes avenues for future research.
\section{Model}
\begin{figure*}[ht]
  \centerline
      {
        \includegraphics[width=.7\textwidth, height=.2\textheight]{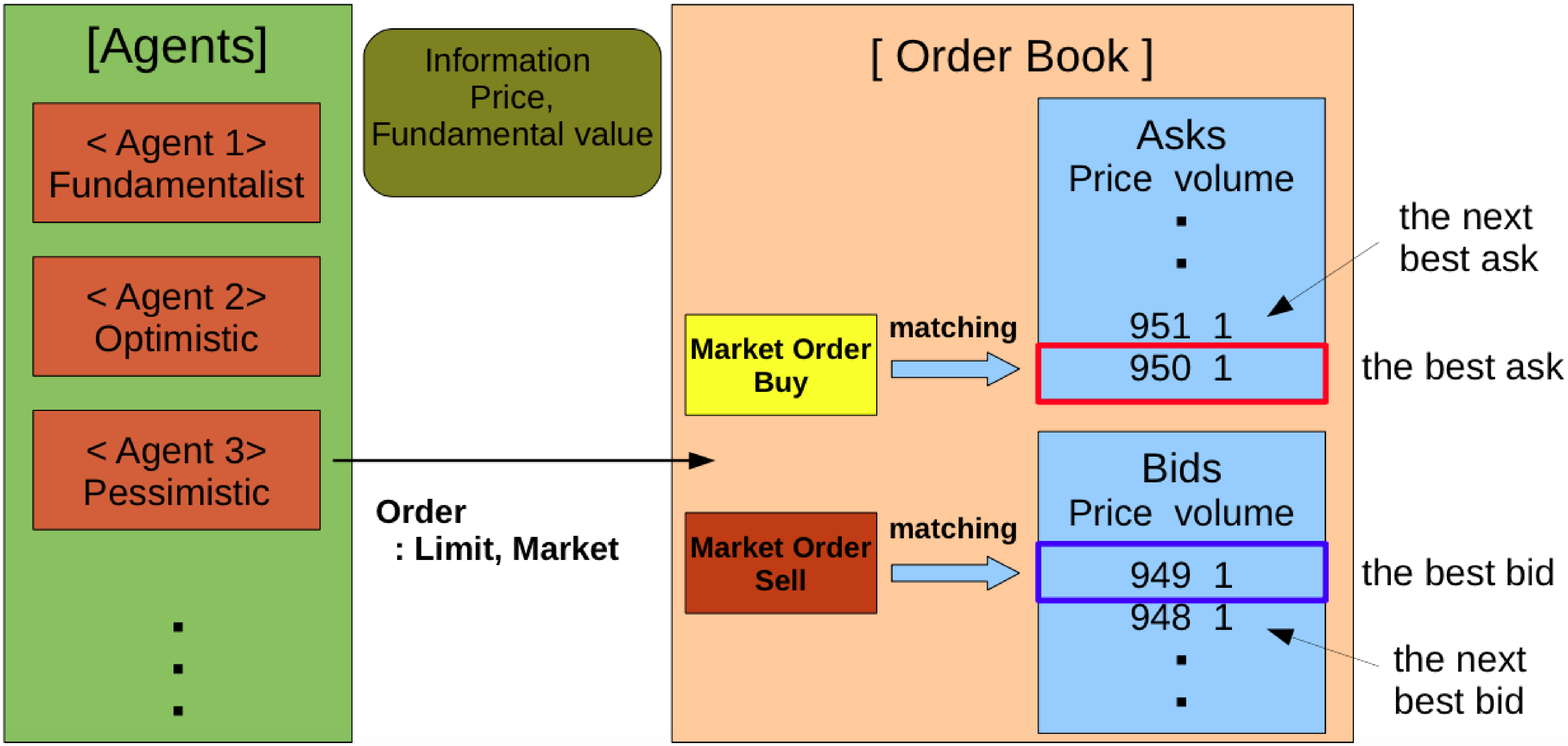}
      }
      \caption{The schema of the trading process incorporated in the ADAM.}
      \label{fig:fig1}
\end{figure*}

Traditional asset pricing models in the financial system are typically characterized by one partial aspect among diverse market states, but the dynamics of asset prices in the real financial system have  many aspects that can be result from the  heterogeneity of traders, the interactions among them, and a complicated trading mechanism (double auction market), which has led us to engage in our study using the ADAM.\\
In the ADAM, it is assumed that one type of stock is traded. We assume that all agents know other agents' type, past price information during their investment time horizon and the current fundamental value $p_{t}^{f}$ of the asset, which we take to follow the geometric Brownian motion defined by \\
\begin{eqnarray}
\ln(p_{t}^{f})-\ln(p_{t-1}^{f}) = \epsilon_{t}, ~\epsilon_{t} \sim N(0,\sigma_{\epsilon})
\label{eq:fundvalue}
\end{eqnarray}
where $p_{t}^{f}$ denotes the fundamental value at time $t$. The increments of $\ln(p_{t}^{f})$ follow a normal distribution that yields a standard deviation equal to $\sigma_{\epsilon}$ for the aggregate of the increments over integer time steps.\\
Fig \ref{fig:fig1} describes the schema of the trading process in the ADAM. At the first step of the trading process, an individual agent determines her own type according to switching rules among agent types. After the process of determining an agent type, one agent randomly chosen forecasts future price and determines her order. Finally, the agent submits the order, which is determined as one type between the market order and the limit order under the double auction mechanism. If an agent submits a market order to buy(sell), this market order is matched with a limit order to sell(buy) at the best ask(bid) price level, and the market price is determined by the best ask(bid) price. If an agent submits a limit order to buy(sell), this order is stored in a bid(ask) limit order book. This trading process can be divided into two sub-processes: the determining agent type process and the double auction market process. The details of these sub-processes are described in the following two subsections.
\subsection{The determining agent type process}
In the ADAM, the trading process begins from determining the agent type. We consider two heterogeneous agent types: the fundamentalist and the chartist. The fundamentalist agent prefers fundamental analysis and corresponds to a fundamental trader in the real financial market \cite{fundamentalist}. In the ADAM, fundamentalists obtains fundamental value information as a result of a fundamental analysis. Generally, the fundamental value of stock is typically calculated by the discounted sum of future profits or the earnings of the company issuing the stock. This fundamental value moves like a random walk and is largely  not affected by the trend of the market price. In the ADAM, fundamental value is assumed to follow a Brownian motion, as represented by Eq.\ref{eq:fundvalue}. Therefore, fundamentalism both stabilizes the market and randomizes market prices. Market states driven by fundamentalists will be similar to the efficient markets proposed by Eugene F. Fama. Unlike fundamentalists, chartists are sensitive to the trends or fashions of market prices. Chartists are agents who prefer technical analyses using past price trends and correspond to technical traders in the real financial market \cite{chartist}. As proposed by Robert J. Shiller, a chartist is an irrational trader and speculator who destabilizes a market. In the ADAM, a chartist is a trend-follower using trends of past prices. Additionally, chartists are classified as two types: optimistic and pessimistic. Optimistic(pessimistic) chartists forecast that future prices will be larger(smaller) than the current price.\\
In most of the previous artificial stock markets using ABMs, both fundamentalists and chartist are mixed in expectations of the spot price form and can be stochastically determined using initial fixed parameters \cite{chiarella1,chiarella2,yamamoto}. These models have the advantage of generating the market state as varying combinations of agent types. However, these models do not provide an adequate description of the dynamical properties of the market microstructure according to a change in the agent's type. To understand the dynamics of the market microstructure according to a change in the agent's type, we segregate the fundamentalists and chartists in expectation form, and we consider switching rules between them using transition probabilities as we modify the rules of opinion dynamics from previous artificial stock market models \cite{lux,lux1}. Transition probabilities include herding that occurs in the interactions between agents and the profit terms of each agent type. More details of switching rules are presented in the Appedix \ref{app:sw_rule}.
\subsection{Double auction market process}
After all agents determine their own types, one agent is randomly chosen to participate in trading at any time $t$. The chosen agent $i$ forms an expectation about the spot price or the future price $\hat{p}^{i}_{t,t+\tau^{i}}$, that will prevail in the time interval $(t,t+\tau^{i})$, where $\tau^{i}$ denotes the investment time horizon of agent $i$. The future prices of fundamentalist, optimistic and pessimistic agents are as follows:
\begin{eqnarray}
\hat{p}^{i}_{t,t+\tau^{i}}(fundamentalist)=p_{t}^{f}(1+N(0,\frac{\sigma_{\epsilon}}{\gamma_{f}}))\\
\hat{p}^{i}_{t,t+\tau^{i}}(optimistic)=p_{t}+|N(0,\frac{\sigma_{\tau^{i}}}{\gamma_{c}})|\\
\hat{p}^{i}_{t,t+\tau^{i}}(pessimistic)=p_{t}-|N(0,\frac{\sigma_{\tau^{i}}}{\gamma_{c}})|\\
\sigma_{\tau^{i}}^{2} = \sum_{k=1}^{\tau^{i}}{({p_{t-k}-\bar{p}})^{2} \sqrt{\tau^{i}}}/\tau^{i}\\
\bar{p} = \frac{\sum_{k=1}^{\tau^{i}}{p_{t-1-k}}}{\tau^{i}}
\end{eqnarray}
where $p_{t}^{f}$ denotes the fundamental value at time $t$, ${\gamma}_{f}$, ${\gamma}_{c}$ denotes the risk aversion coefficient of fundamentalists and chartists, and $\tau^{i}$ denotes the investment time horizon of agent $i$. We assume that ${\gamma}_{f}$ is larger than ${\gamma}_{c}$. Also,we assume that $\tau^{i}$ of fundamentalists is larger than that of chartists. These assumptions reflect the characteristics that chartists are more speculative than fundamentalists. $N(0,\frac{\sigma_{\tau^{i}}}{\gamma_{c}})$ is normal distribution with a mean of zero and $\frac{\sigma_{\tau^{i}}}{\gamma_{c}}$, as the standard deviation. $\sigma_{\tau^{i}}$ denotes the standard deviation of prices during $[t-\tau^{i},t)$. Chartists take the standard deviation of future prices from the market data. Optimistic(pessimistic) agents drive the market price to up-trend(down-trend) with the standard deviation of the market price. Fundamentalists take the standard deviation of the future price with $\sigma_{\epsilon}$, which is the standard deviation of the increments of fundamental values.\\
 If the agent expects that the future price will be larger (smaller) than current price, she decides to buy(sell) one unit of the stock. However, if the agent expects that the future price is the same as the current price, the agent does not submit an order. We assume that the agent is willing to buy (sell) at a price $b_{t}^{i}(a_{t}^{i})$ that is lower(higher) than his expected future price $\hat{p}^{i}_{t,t+\tau^{i}}$. $b_{t}^{i}$ and $a_{t}^{i}$ are as follows:
\begin{eqnarray}
b_{t}^{i} = \hat{p}_{t,t+\tau^{i}}^{i}(1-k^{i}(x))\\
a_{t}^{i} = \hat{p}_{t,t+\tau^{i}}^{i}(1+k^{i}(x))\\
Prob(k^{i}(x))= \exp{(-x/\sigma)}/\sigma \\
E[k^{i}(x)]=\sigma~~, Var[k^{i}(x)] = \sigma
\end{eqnarray}
where $k^{i}$ is distributed by an exponential distribution. According to the double auction mechanism, the agent selects a limit order or market order. If $b_{t}^{i}$($a_{t}^{i}$) is smaller (larger) than the best ask $a_{t}^{q}$(the best bid $b_{t}^{q}$), the agent submits a limit order at the price level $b_{t}^{i}$($a_{t}^{i}$). The best ask $a_{t}^{q}$(the best bid $b_{t}^{q}$) means that the lowest ask(the highest bid) is listed in a limit order book at time t. A limit order is an order to buy or sell a stock at a specific price or better and is stored in an order book that waits for market orders. However, if $b_{t}^{i}$($a_{t}^{i}$) is larger (smaller) than or equal to $a_{t}^{q}$($b_{t}^{q}$), the agent submits a market order at $a_{t}^{q}$($b_{t}^{q}$). A market order is an order to buy or sell a stock at the best available price in a limit order book. When a market order is submitted in the market, a transaction occurs. Until $t + \tau^{i}$, limit orders that are unmatched with a market order are removed from the order book.\\
 At any time, $t$ the price $p_{t}$ is given by the price at which a transaction, if any, occurs. If no new transaction occur, a proxy for the price is given by the average of $a_{t}^{q}$ and $b_{t}^{q}$ so that $p_{t} = (a_{t}^{q} + b_{t}^{q})/2$, which is a value that we call the mid-point. If no bid or ask price is listed in the order book, a proxy for the price is given by the previously traded or quoted price. All prices of orders must be positive, and investors can submit limit orders at any price on a prespecified grid, as defined by the tick size $\Delta$.\\
 All agents trade under a budget constraint, and short sales are forbidden. To prevent an overly large change in market price, a submission of an order that exceeded or was below 15 \% of the closed price at a previous time is forbidden. We assume that the volume of submitted orders is always one unit, and one simulation step, denoted by $\Delta t$ is 0.01 time. We assume one trading period is 100 simulation steps (=100$\Delta t$ = 1 time).\\
 After trading is completed, all limit orders and market microstructure trajectories are written in artificial data. By analyzing of these artificial data, we can trace all the processes of price formation at a micro-level.\\
\section{Results}
We employ the ADAM model developed above to understand the asset pricing model on a market microstructure level. According to the numerous results from previous works, the asset price dynamics in microstructure scope are characterized by several quantities, including the return, volatility, bid-ask spread, and first gap data sets \cite{zovko,bouchaudorder,potters,becker,flannery,carlson,plerou2,farmer,lillo,weber,gillemot,ponzi}. Here, we simulate the artificial data sets using the ADAM model, which can generate significant information for understanding asset price mechanisms, particularly in terms of the heterogeneity of traders. We now proceed with the simulation and analyze the data model generated. A total of 100 simulations are performed with a different random seed. Each simulation is performed with 500 agents in 1,000,000 simulation steps(=10,000 times). The values of the parameters used for simulations are as follows:\\
$\sigma_{\epsilon}$ = 0.005, $\sigma$ = 0.1, $\Delta$ = 0.0005, $p(t=0)$ = 300, $p_{f}(t=0)$ = 300,${\gamma}_{f}$ = 1.0, ${\gamma}_{c}$ = 0.1, and $\tau^{i}$ = 3 [time] for fundamentalists; and $\tau^{i}$ = 1 [time] for chartists. To protect market distortion in the ADAM model, we construct the marker collapse state to recover to other states by virtue of the rule which forbids trading in excess or below of 15\% of the market price in previous trading periods.\\
Fig. 2(a) depicts a large fluctuation over time, which indicates that an intermittent increase in the number of chartists tends to deviate from that of the efficient market, which consists only of fundamentalists. Deviation from this behavior, however, is an efficient market hypothesis. Fig. 2(b) - (d) shows the temporal movement for the return time series, volatility, the bid-ask spread, and the first gap, respectively. The bid-ask spread is calculated by the difference between the best ask and the best bid price, and the first gap also corresponds to the difference between the best bid(ask) and the next best bid(ask). In Fig. 2 (b)-(d), we find that  the temporal movement of the four variables should be caused by the ratio of chartists in the market, which is a finding supported previously by both theoritical and ABMs \cite{boswijk,hommes2,chiarella1}. Based on the results from Fig. 2, the degree of information asymmetry induced from the chartist traders plays an important role in the transition behavior among diverse market states. In the Appendix \ref{app:facts}, we also calculate the statistical and dynamical properties for four variables using the detrended fluctuation analysis and the probability density function; similar to the results from previous work \cite{plerou2,farmer,lillo,weber,gillemot,ponzi,liu,li,ag}, we find that the long-range correlation observed in all data sets, except for the return time series and the distribution function, follows a power-law distribution function regardless of data sets. We verify the usefulness of the proposed model by analyzing the time series properties of the artificial data sets generated by the ADAM.\\
Considering the significant role that the heterogeneous investors in the financial market play in creating the complicated market structure against the EMH market with respect to homogeneous traders, it is natural to expect that the degree of heterogeneity for traders is essential to generate diversity in financial makret conditions. To investigate the validity of this hypothesis in terms of understanding the fundamental feature of asset prices in the diverse market states, we analyze whether the various market states, such as normal, abnormal, and collapsed are determined by the ratio of chartists as the source of abnormal market condition. To analyze the relationship between the heterogeneity of traders and market states, we divide the market microstructure data of the ADAM model into a sub-data set based on the ratio of chartist traders, $P_c$. As Fig. \ref{fig:fig3}(a),(c),(e) shows, for three quantities volatility, bid-ask spread, and first gap the ratio of chartist traders, $P_c$, is an important factor that describes diverse market states. In Fig. \ref{fig:fig3} (a),(c),(e), we depict the probability of extreme events, $N_e$ and the volume of limit order books (LOB), as a function of the $P_c$
, of those data sets. $N_e$ is measured by the event exceeded by four standard deviations, $4\sigma$. In all cases, $N_e$ and the volume of LOB is significantly related to $P_c$, which indicates that in three quantities, $P_c$ plays an important role in terms of determining market states. In Fig. \ref{fig:fig3}(a),(c),(e), we find that, regardless of data set, three distinct market states are observed and define each market state as mimicking efficient market hypothesis(MEMH, $0 \leq P_c^{\text{volatility}} \leq 0.45$, $0 \leq P_c^{\text{bid-ask spread}} \leq 0.44$, and $0 \leq P_c^{\text{first-gap}} \leq 0.4$), mimicking real financial market(MRFM, $0.46 \leq P_c^{\text{volatility}} \leq 0.84$, $0.45 \leq P_c^{\text{bid-ask spread}} \leq 0.84$, and $0.41 \leq P_c^{\text{first-gap}} \leq 0.84$), and mimicking market collapse(MMC, $P_c \geq 0.85$) states. MRFM states are defined by a threshold value larger than $N_e = 0.005$. These results show that the MEMH state is similar to the efficient market proposed by Fama \cite{fama}. In an MEMH state, fundamentalists are more prevalent than chartists in the market. With a substantial number of fundamentalists, the market price converges to a fundamemental value and reflects full rationality. The volume of LOB increases because of the vastly limit orders submitted by fundamentalists in the MEMH state, indicating that fundamentalists can play a role in the liquidity provider. An MEMH state in the specific range of $P_{c}$ implies that the efficient market is only an ideal state or one type of states(out of a variety of possible states) in the financial market. In an MEMH state, the limit order volume decreases with increases of $P_{c}$ (Fig. \ref{fig:fig3}(a)(c)(e)), which implies that most market order submissions or transactions are perfomred by chartists. For an MEMH state in which the fundamentalists are more dominant than chartists in the market, the probability density function (PDF) of those variables follows a Gaussian distribution function, which is similar to the EMH proposed by Eugene F. Fama \cite{fama} and indicates that the market price converges to a fundamental value that should be generated by fully rational agents based on complete information. The essential features of market prices generated only by fundamentalists is similar to that of EMH (Appendix \ref{app:hree}). For an MRFM state, there is a positive correlation between the $N_e$ of three variables and the $P_c$ value, and the PDF in Fig.\ref{fig:fig3}(b)(d)(f) follows a power-law distribution that is observed in a real financial market, which reveals that the increasing number of chartists generates extreme events that are frequently shown in the real financial market. This finding indicates that the prices in an MRFM state have trends or fashions and deviate from efficient market states, as proposed by Robert J. Shiller \cite{shiller}. In other words, because there are more chartists than fundamentalists in the market, the market price becomes largely separated from fundamental value (random walk).
 We also find that the volume of limit order book (LOB) and $P_c$ shows a negative relationship in Fig. \ref{fig:fig3}(a),(c),(e). 
In an MMC state, we find that $N_e$ drops to zero due to the shortage of transactions, and the volume of LOB is almost zero. According to these results, the chartists should behave as liquidity takers and can sharply increase the liquidity risk. An MMC state created by chartists will ultimately lead to a market collapse because of the liquidity shortage. \\ 
To analyze the characteristics of limit order book for three distinct market states, we show snapshots of the LOB of each state. Fig. 4 (a),(b),(c) shows the LOB of the MEMH($P_c=0.08$), MRFM($P_c=0.62$), and MMC($P_c=0.85$) states, respectively. In Fig. 4 (a),(b),(c), we find that the bid-ask spread and width between limit orders increase as $P_c$ increases; thus, it is the smallest in the MEMH state and has the largest value in the MMC state that is characterized by a liquidity shortage or liquidity evaporation. In Fig. 4(b), we find that the suitable ratio of chartists increases the liquidity of transactions among agents and that it will reduce the liquidity risk. Thus, the stability of the financial market should be controlled by the ratio of chartists in the entire population of traders. Previous empirical studies have reported that the bid-ask spread in a financial crisis is larger than it is before the crisis \cite{becker,flannery}. It is reported that specialists in the NYSE (New York Stock Exchange) had trouble executing trades because of the dramatic shortage of limit orders on Black Monday on Octtober 19,1987 \cite{carlson}. Our model is consistent with these empirical findings, including the liquidity shortage on Black Monday.\\
 The standard deviation of volatility, the bid-ask spread and the first gap as a function of $P_{c}$ increase and have peak points at approximately 0.85 (Fig. \ref{fig:fig5}). All maximum standard deviation values of these quantities are normalized by 1. When $P_{c}$ is greater than 0.85, the standard deviation of the three quantities dramatically decreases and is close to zero in Fig. \ref{fig:fig5}. 
Our results suggest that the behavior of LOB seems like a transition from an alive state to a collapsed state at around $P_c=0.85$, which is similar to the phase transition behavior described by previous research \cite{stanley}.\\
We observe various stylized facts in asset returns and market microstructures such as power-law tails, a long memory of volatility, first gap and aggregational Gaussianity \cite{plerou2,farmer,lillo,weber,gillemot,ponzi,liu,li,ag}. To confirm the effect of the chartists, we performe an analysis of a homogeneous equilibrium market that consistes of only fundamentalists. As a result, market prices converge to the fundamental value, and the market state does not deviate from the MEMH state during simulation. MRFM, MMC and stylized facts are not observed in the homogeneous equilibrium market. The details of analyzing stylized facts and the results of the homogeneous equilibrium market are can be found in the Appedix \ref{app:facts}, \ref{app:hree}.\\ 

 \section{Discussion and conclusion}
Understanding asset pricing is a crucial issue in the financial market; however, because there is no adequate theory in finance to explain it that includes all aspects of the price mechanism, most theory reflects the partial features of the manner in which asset prices are determined by the investment strategies of agents. \\
Indeed, applying important elements, including the heterogeneity of traders and the double auction market as trading system, to an artificial market should consider a diversity of market states. In this study, we have suggested the ADAM model to illustrate the diversity of the financial markets from an efficient market to behavioral finance. The ADAM model consist of a heterogeneity of traders, including fundamentalists and chartists, and provides a differentiated framework compared with previous ABMs. Our key finding is that the diversity of the financial market, including the MEMH, MRFM, and MMC, is determined primarily by the ratio of chartists, $P_c$, among all the traders in the market. \\
The ADAM framework developed herein offers explanations and raises questions that can further deepen our understanding of asset pricing models in financial markets. For example, although the ADAM cannot be considered a multi-asset pricing model, the framework we develop can explain the diversity of financial market states and can provide a framework with which to address the role of the heterogeneity of agents. \\
In summary, our findings argue that many aspects of market states can be explored numerically by the ABM if we combine the heterogeneity of traders and the double auction market as trading systems, thus opening a new route to a more in-depth understanding of financial markets. \\
\appendix

\section{\label{app:sw_rule} The switching rules of agents' opinions}
 In this paper, we apply the modified transition rules of agents' opinions which are based on Lux and Marchesi(1999) \cite{lux}. The transition probabilities introduced in Lux and Marchesi(1999) \cite{lux} denoted by $\pi_{A,B} \Delta t$. $\pi_{A,B}$ is the transition rate from type B to type A.\\
\begin{eqnarray}
\nonumber \pi_{+-} = v_{1}\frac{n_{c}}{N}\exp{(U_{1})},\pi_{-+}=v_{1}\frac{n_{c}}{N}\exp{(-U_{1})},\\
\nonumber U_{1} = \alpha_{1}x + \frac{\alpha_{2}}{v_{1}}\frac{dp/dt}{p}\\
\nonumber \pi_{+f}=v_{2}\frac{n_{+}}{N}\exp(U_{2,1}),\pi_{f+}=v_{2}\frac{n_{f}}{N}\exp(-U_{2,1})\\
\nonumber \pi_{-f}=v_{2}\frac{n_{-}}{N}\exp(U_{2,2}),\pi_{f-}=v_{2}\frac{n_{f}}{N}\exp(-U_{2,2})\\
 \nonumber U_{2,1}=\alpha_{3}\{\underbrace{\frac{r+(1/v_{2})(dp/dt)}{p}-R}_{\text{profit of optimistic}}-\underbrace{s|\frac{p_{f}-p}{p}|}_{\text{fundamentalists' profit}}\}\\
\nonumber U_{2,2}=\alpha_{3}\{\underbrace{R-\frac{r+(1/v_{2})(dp/dt)}{p}}_{\text{profit of pessimistic}}-\underbrace{s|\frac{p_{f}-p}{p}|}_{\text{fundamentalists' profit}}\}
\end{eqnarray}
where $x = (n_{+}-n_{-})/n_{c}$,$N,n_{c},n_{+},n_{-},n_{f}$ denote the number of agents, chartists, optimistic agents, pessimistic agents, and fundamentalists.$p$ denotes the current market price, and $p_{f}$ is the current fundamental value. The subscript $+,-,f$ denotes the agent type, e.g., optimist, pessimist, or fundamentalist. The mechanism that can change the agent types in transition probability consists of a herding and profit strategy. If the fraction of optimisticagents ($n_{+}/N$) which is related to herding strategy increases in the market, the transition probabilities from other types to optimistic agents($\pi_{+-},\pi_{+f}$) increase. In addition, if the profit of opimistic agents increases, the transition probabilities from other types to optimistic agents increases. The other types such as pessimists and fundamentalists also have similar behaviors.\\
To avoid absorbing state($n_{c}$ = 0 or $n_{f}$ = 0) in simulation, we apply a rule that an agent with an opinion in a population less than 0.8\% in the total population can not change her own opinion into other opinions.\\
We assume that the investment time horizon of each agent type is different. Chartists have shorter investment time horizon than fundamentalists. However, in the transition probabilities in Lux and Marchesi(1999) \cite{lux}, the heterogeneous investment time horizon of an agent is not considered. In other words, $dp/dt$ is calculated during homogeneous investment time horizons. However, in the real financial market, investors have heterogeneous information sets. Each investor cannot help but have a different strategies due to the differences between information sets that agents have. With the notion of heterogeneous information sets between agents, we modify this term $dp/dt$, which can reflect the investment time horizon of heterogeneous agent types. $dp/dt$ is calculated by the average value of $\Delta p/\Delta t$ during agents' own investment time horizon [$t-\tau^{i}$,t).\\
The values of the parameters are set as follows: N (number of agents) = 500, $v_{1}$ = 2, $v_{2}$ = 0.6, $\alpha_{1}$ = 0.6, $\alpha_{2}$ = 1.5, $\alpha_{3}$ = 1, $r$ = $Rp_{f}(R = 0.0004)$, $s$ = 0.75, $\Delta t$ = 0.01 [time] ,$\tau^{i}$ = 3 [time] investment time horizon for fundamentalists, $\tau^{i}$ = 1 [time] for chartists.\\
For more details about other parameters in transition probabilities, see Lux and Marchesi(1999) \cite{lux}.\\
\section{\label{app:facts} Stylized facts in Artificial Double Auction Market}
 We analyze ADAM-generated data and observe various `stylized facts' that were observed in previous empirical studies \cite{plerou2,farmer,lillo,weber,gillemot,ponzi,liu,li,ag}. Fat-tails of market microstructure quantities such as absolute return, bid-ask spread and first gap, which were widely observed in previous empirical studies, are observed in the ADAM(Fig. \ref{fig:figA1}). We fit tails of the CDF (Cumulative Distribution Function) of market microstructure quantities using a power-law function, $y \sim x^{-\alpha}$ and $\alpha$ is estimated by maximum likelihood estimation \cite{newman}. The fitted results are summarized in Table \ref{tab:tabA1}. Additionally, the distribution of return as time lag increases converges to the return distribution of fundamental value, which follows a normal distribution(Fig. \ref{fig:figA2}(a)), which is called `Aggregational Gaussianity', which is one type of `stylized fact' in the financial markets \cite{ag}.\\
To investigate the temporal correlation property of market microstructure, we use the DFA (Detrended Fluctuation Analysis) method introduced in Hu et al (2001) \cite{kunhu}. The steps of the DFA are as follows:\\
(i) We consider time series $x(i)$ ($i=1,...,N_{max}$) noise induced. We integrate the time series $x(i)$, and which is divided into boxes of equal size $n$.\\
\begin{eqnarray}
  y(j)=\sum_{i=1}^{j}[x(i)-<x>]
\end{eqnarray}
where \\
\begin{eqnarray}
  <x>=\frac{1}{N_{max}}\sum_{j=1}^{N_{max}}x_{i}
\end{eqnarray}
(ii) In each box, the integrated time series $y(i)$ is fitted by a polynomial function, $y_{fit}(i)$, which is called a local trend. For order-$l$ DFA(DFA-1 if l=1, DFA-2 if l=2, etc.), we could apply the $l$-order polynomial function for the fitting. Time series $y(i)$ is detrended by subtracting the local trend $y_{fit}(i)$ in each box, and we calculate the detrended fluctuation $Y(i)$.\\
\begin{eqnarray}
Y(i) = y(i)-y_{fit}(i)
\end{eqnarray}
For a given box size $n$, the root mean square(rms) fluctuation function $F(n)$ is calculated as\\
\begin{eqnarray}
  F(n) = \sqrt{\frac{1}{N}\sum_{i=1}^{N_{max}}{[Y(i)]}^{2}}
\end{eqnarray}
(iii) We repeat the above computation for box sizes $n$ to find a relationship between $F(n)$ and $n$, $F(n) \sim {n}^{H}$.\\
This power-law relationship indicates the presence of scaling between $F(n)$ and the box size $n$. The parameter $H$ is called the Hurst exponent or the scaling exponent and represents the correlation property of the signal: If $H=0.5$,the signal has no correlation (white noise); if $H < 0.5$, the signal has a mean-reverting property or a short memory; if $H > 0.5$, the signal has a persistent property or a long memory. In this paper, DFA-1 is used.\\
 As a result, long memory is observed in volatility, the bid-ask spread, the volume and the first gap in the ADAM. In the return case, no memory is observed. These memory properties of the ADAM support previous previous empirical works \cite{plerou2,lillo,liu,oh}. Fig. \ref{fig:figA3}(a),(b) depicts the rms fluctuation functions $F(n)$ of these quantities. The results of the ADAM and previous empirical results are summarized in Table \ref{tab:tabA2}.
\section{\label{app:hree} The case of the homogeneous equilibrium market}
To distinguish the effect of chartists in the market, we investigate the homogeneous equilibrium market, which consists only of fundamentalists (i.e. there are no chartists in the market.) As Fig. \ref{fig:figA4}shows, there is no volatility clustering or a significant peak of the bid-ask spread or the first gap. These results imply that there is only an MEMH state in the homogeneous equilibrium market. There are no fat-tails in the CDF of absolute return, the bid-ask spread and the first gap (Fig. \ref{fig:figA2} (b), Fig. \ref{fig:figA5}). We measure the Hurst exponent using DFA-1. Fig. \ref{fig:figA3} (c) and (d) depict the rms fluctuation $F(n)$ of the given market microstructure time series. The Hurst exponents of these quantities are similar to 0.5 (Table \ref{tab:tabA3}), which indicates that there is no memory property in the homogeneous equilibrium market.\\
 In short, when all agents are fundamentalists, abnormal market behavior, such as financial crisis does not occur. In additional, `stylized facts' in the real financial market, such as long memory, fat-tails  and Aggregational Gaussianity are not observed.\\
These results imply that chartists are the source of abnormal behavior in the ADAM.

\begin{acknowledgments}
  This work was supported by the National Research Foundation of Korea(NRF) grant funded by the Korea government(MEST) (No.2013017095) and by the reserach funds from Chosun University,Korea,2015.
\end{acknowledgments}

\bibliography{Arxiv_Submit}

\begin{figure*}[t]
  \centerline
  {
    \includegraphics[width=1.0\textwidth, height=.7\textheight]{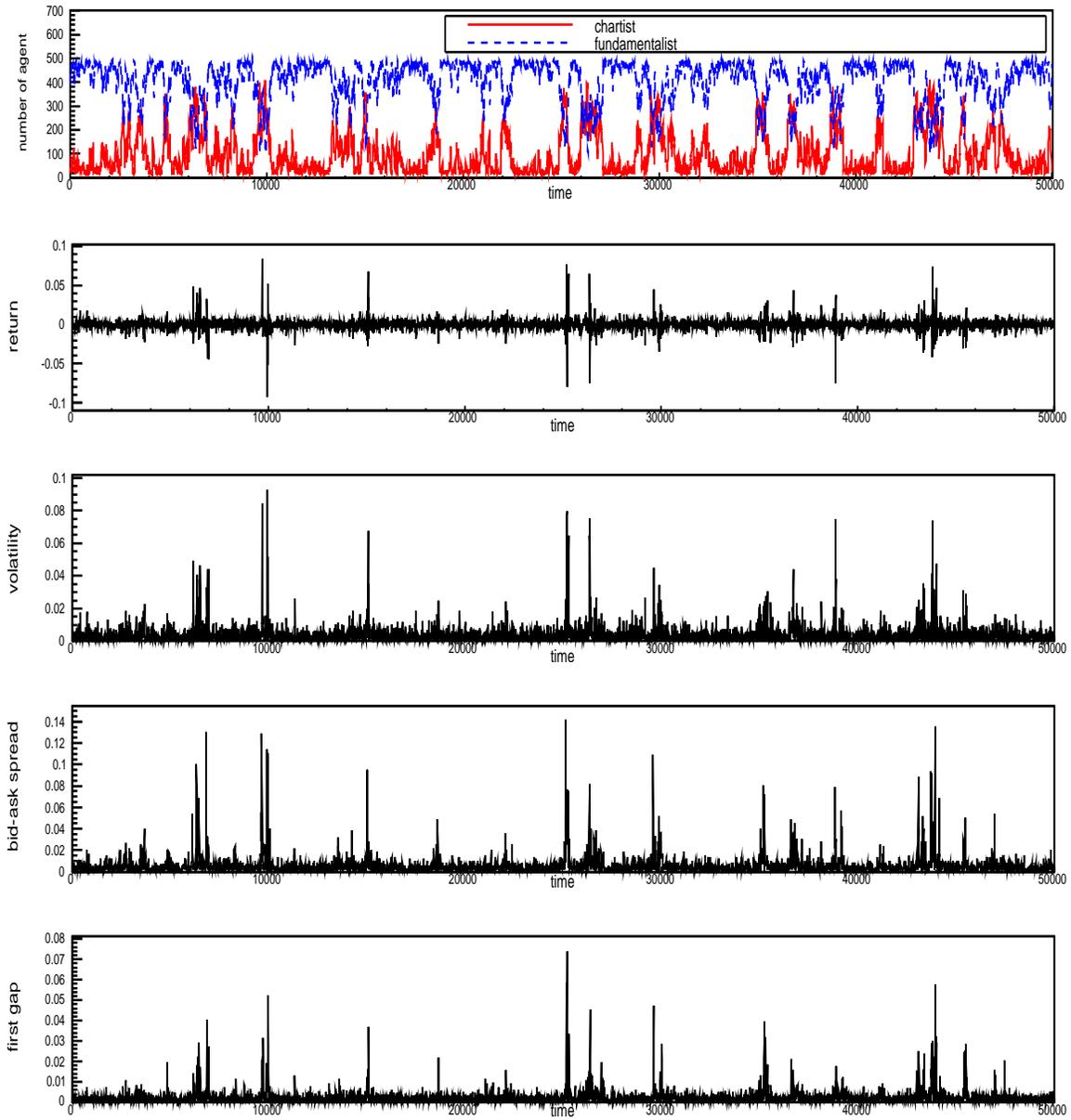}
  }
  \caption{From the top to bottom in figures, the dynamics of the type of agent in the makret, return, the volatility, the bid-ask spread and the first gap as a function of time. In the top figure, the solid red line represents the number of chartists,  and the dashed blue line represents the number of fundamentalists in the market.}
  \label{fig:fig2}
\end{figure*}

\begin{figure*}[ht]
  \centerline
  {
    \includegraphics[width=.5\textwidth, height=.2\textheight]{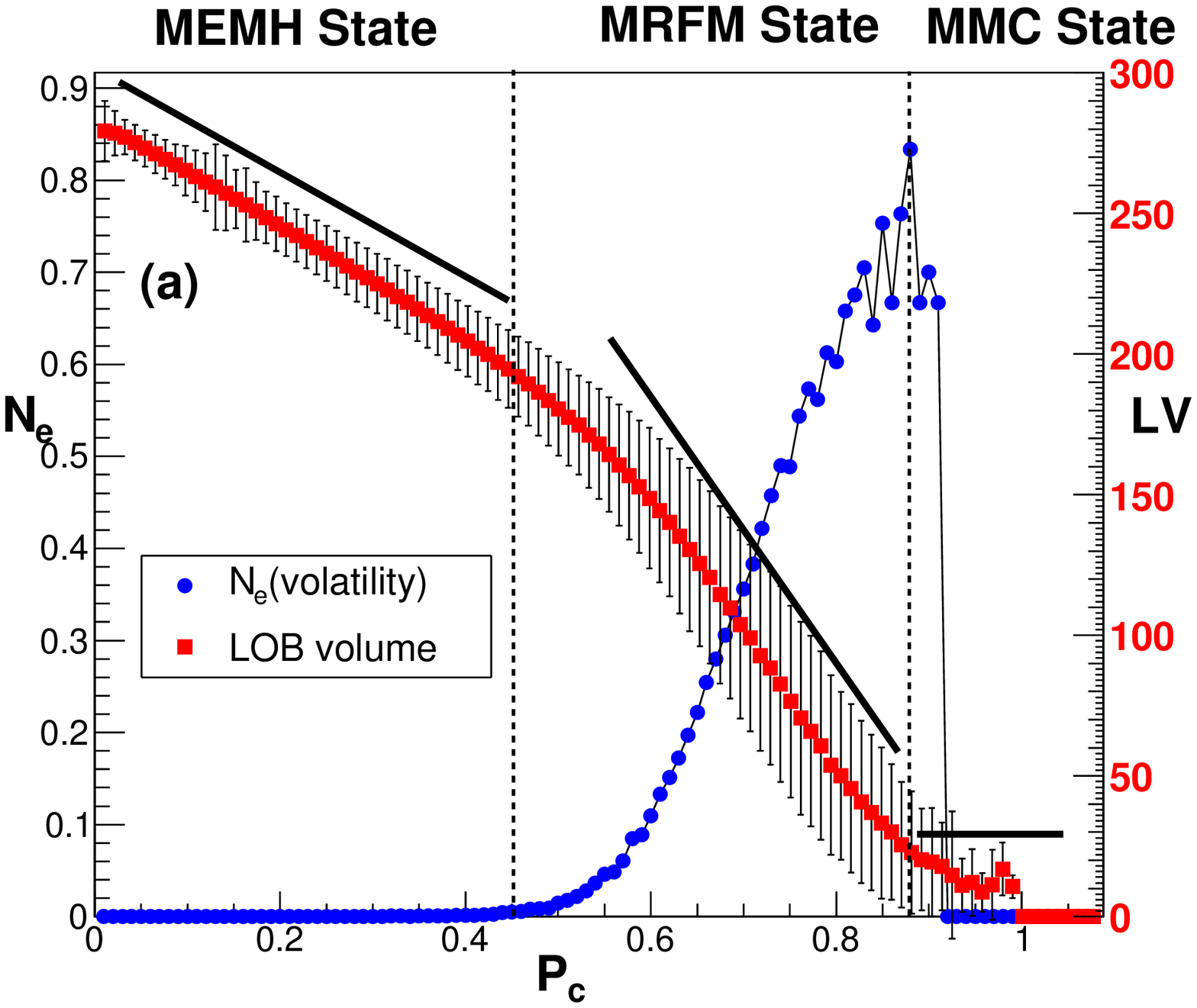}
    \includegraphics[width=.5\textwidth, height=.2\textheight]{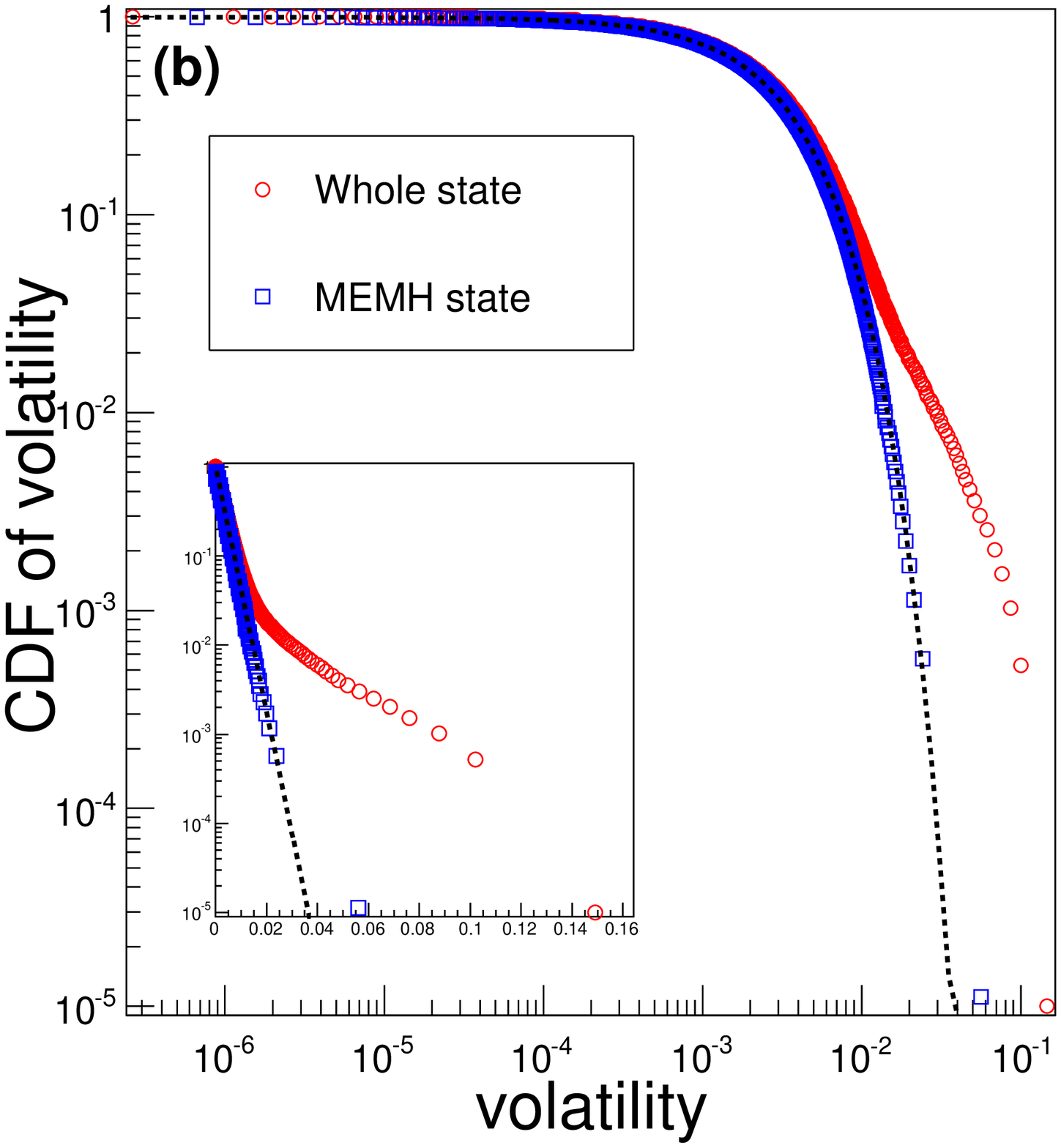}
  }
  \centerline
  {
    \includegraphics[width=.5\textwidth, height=.2\textheight]{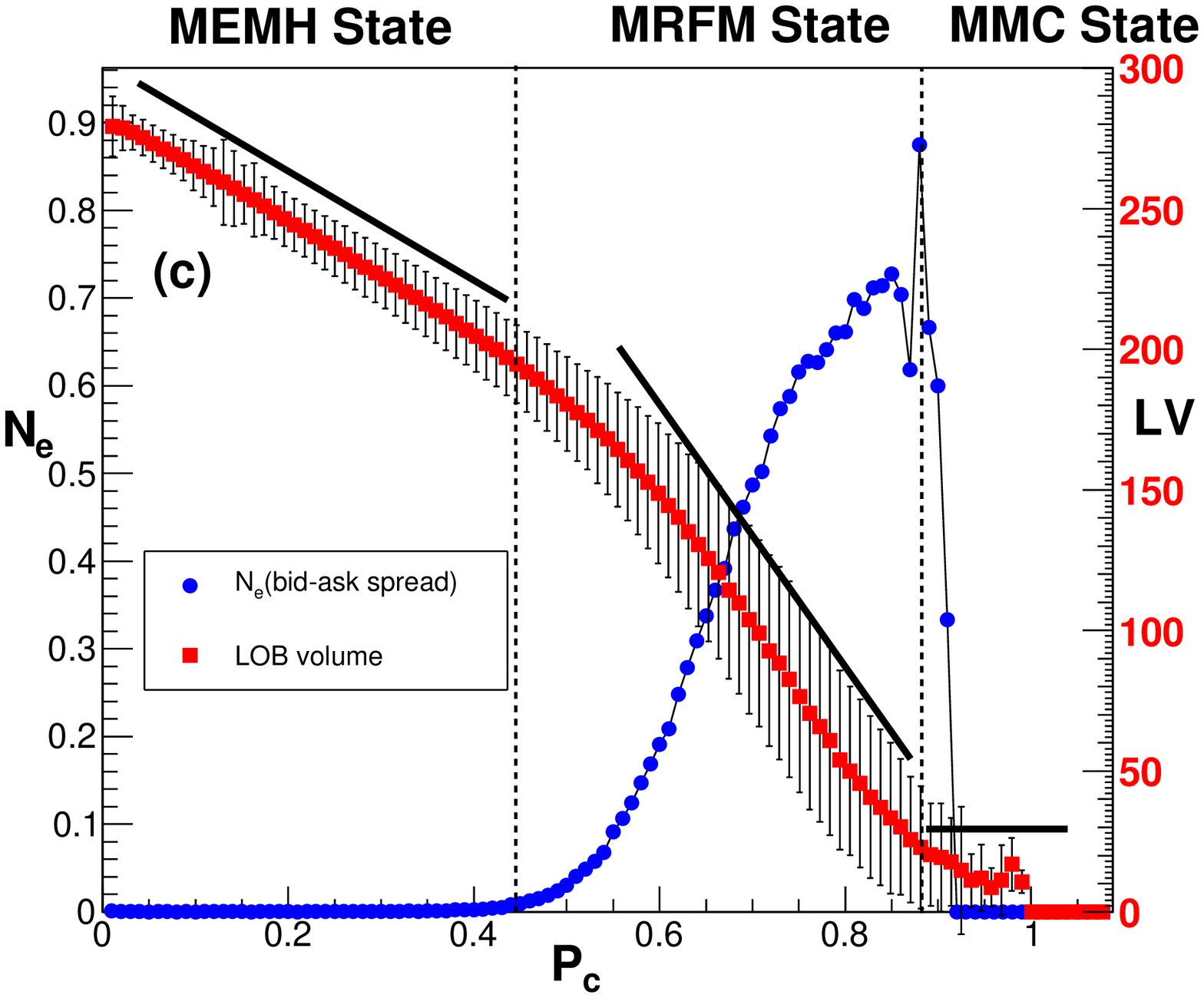}
    \includegraphics[width=.5\textwidth, height=.2\textheight]{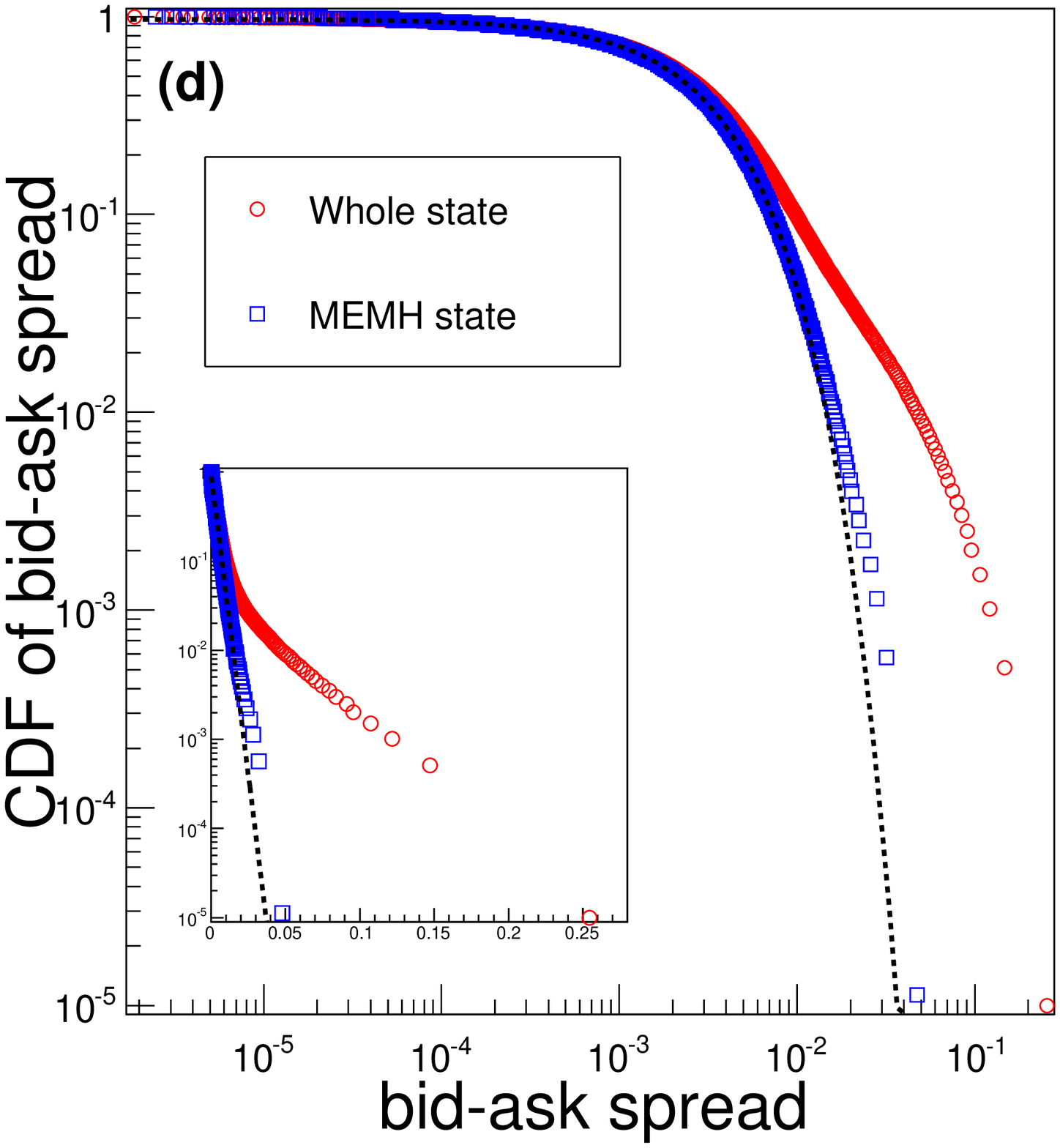}
  }
  \centerline
  {
    \includegraphics[width=.5\textwidth, height=.2\textheight]{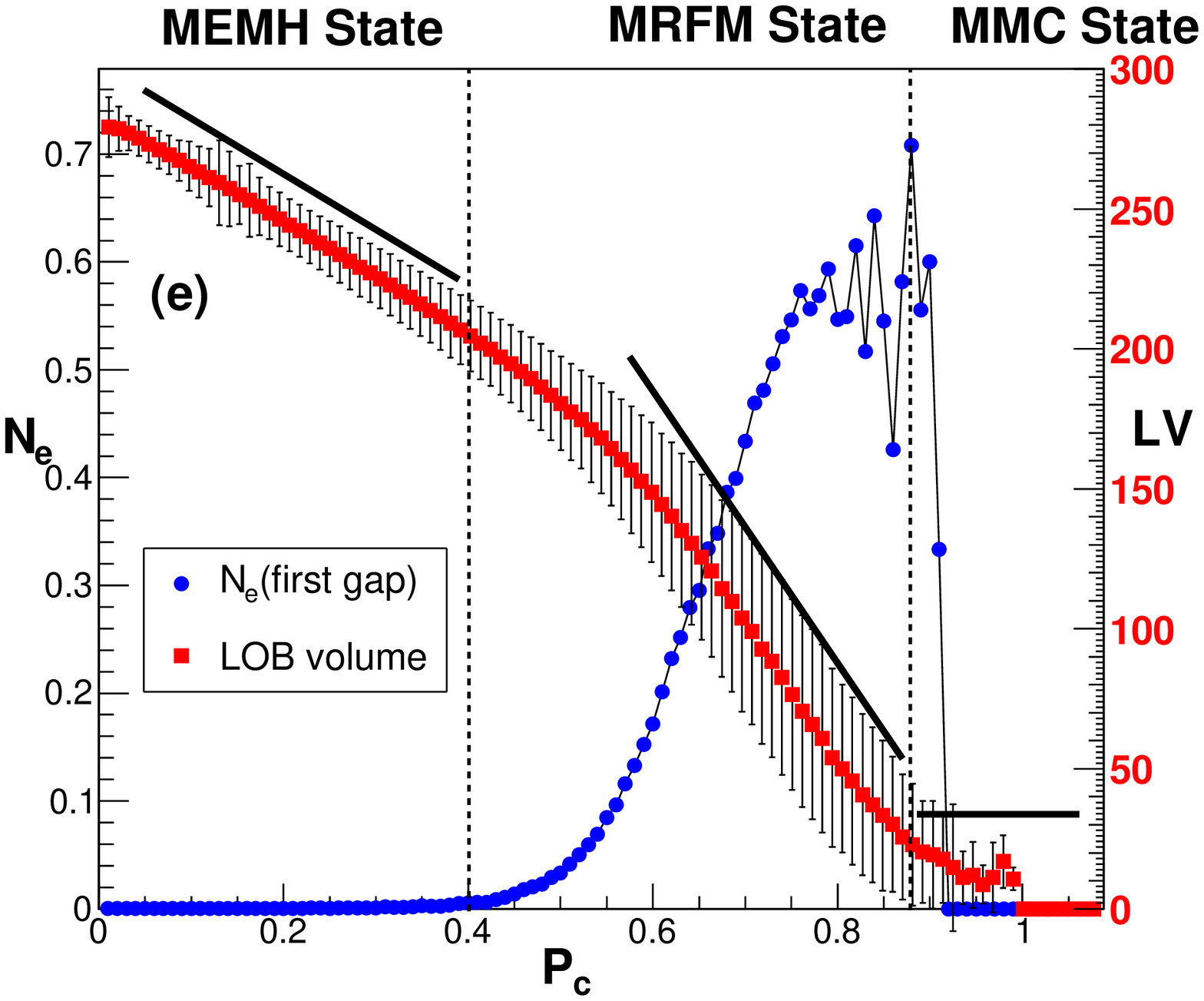}
    \includegraphics[width=.5\textwidth, height=.2\textheight]{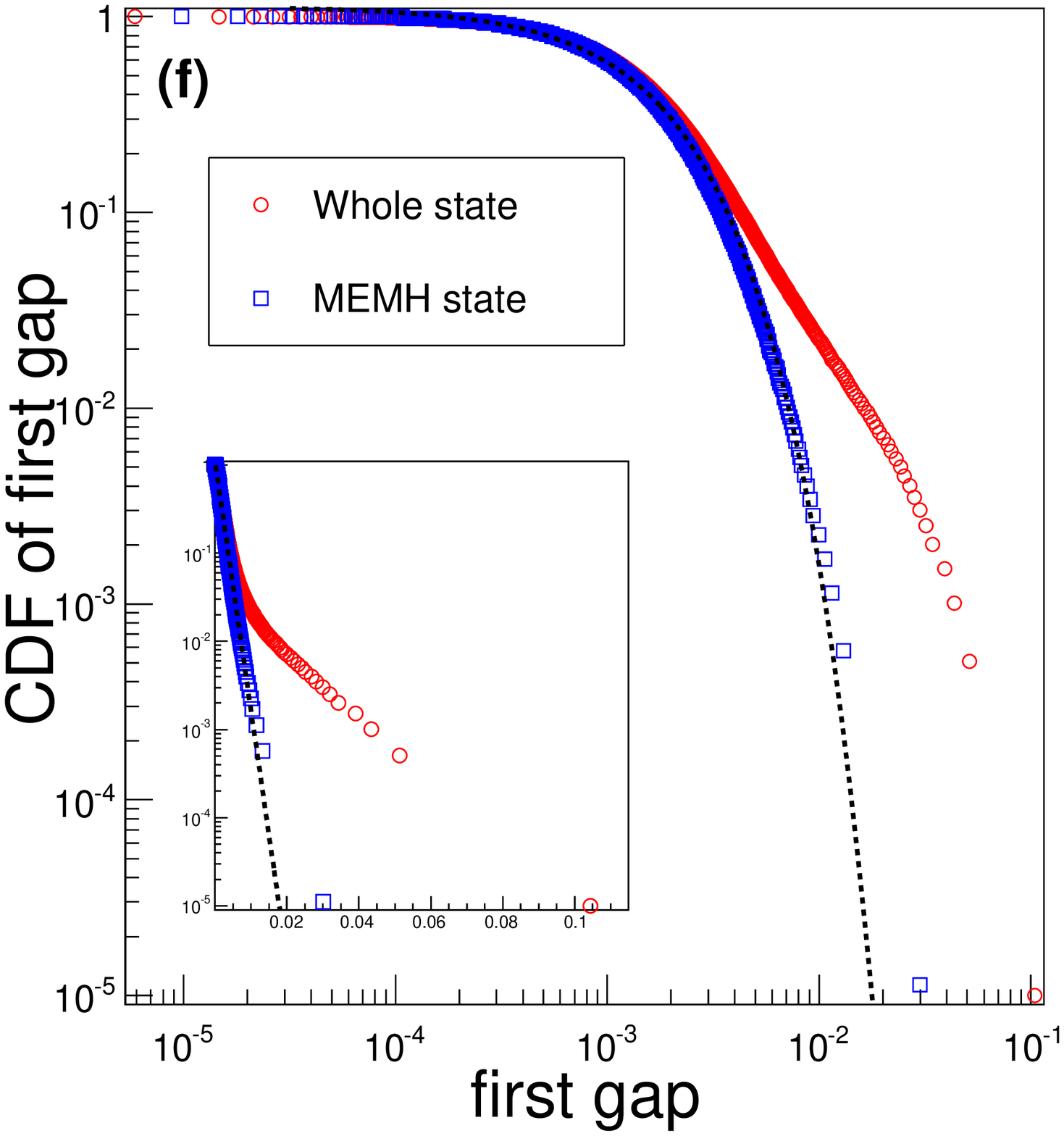}
  }
  \caption{(a)(c)(e) There are $N_{e}$ and LV(limit order book volume) as a function of $P_{c}$. $N_{e}$ denotes the number of extreme events normalized by the number of total events. We define a financial crisis event whose value is larger than 4 $\sigma$. $\sigma$ is the standard deviation of the given time series. A blue filled circle depicts $N_{e}$, and a red filled square depicts the limit order book volume. According to $P_{c}$, we classify the market into three distinct states:  MEMH(Mimicking Efficient Market Hypothesis), MRFM(Mimicking Real Financial Market), or MMC(Mimicking Market Collapse). (b)(d)(f) There are CDF(Cumulative Distribution Function) of volatility, the bid-ask spread and the first gap in an MEMH state and the entire state. A red blank circle represents the CDF of an entire state. The blue-black square represents the CDF of an MEMH state. A dashed black line represents an exponential distribution function.}
  \label{fig:fig3}
\end{figure*}

\begin{figure*}[ht]
  \centerline
  {
    \includegraphics[width=.75\textwidth, height=.25\textheight]{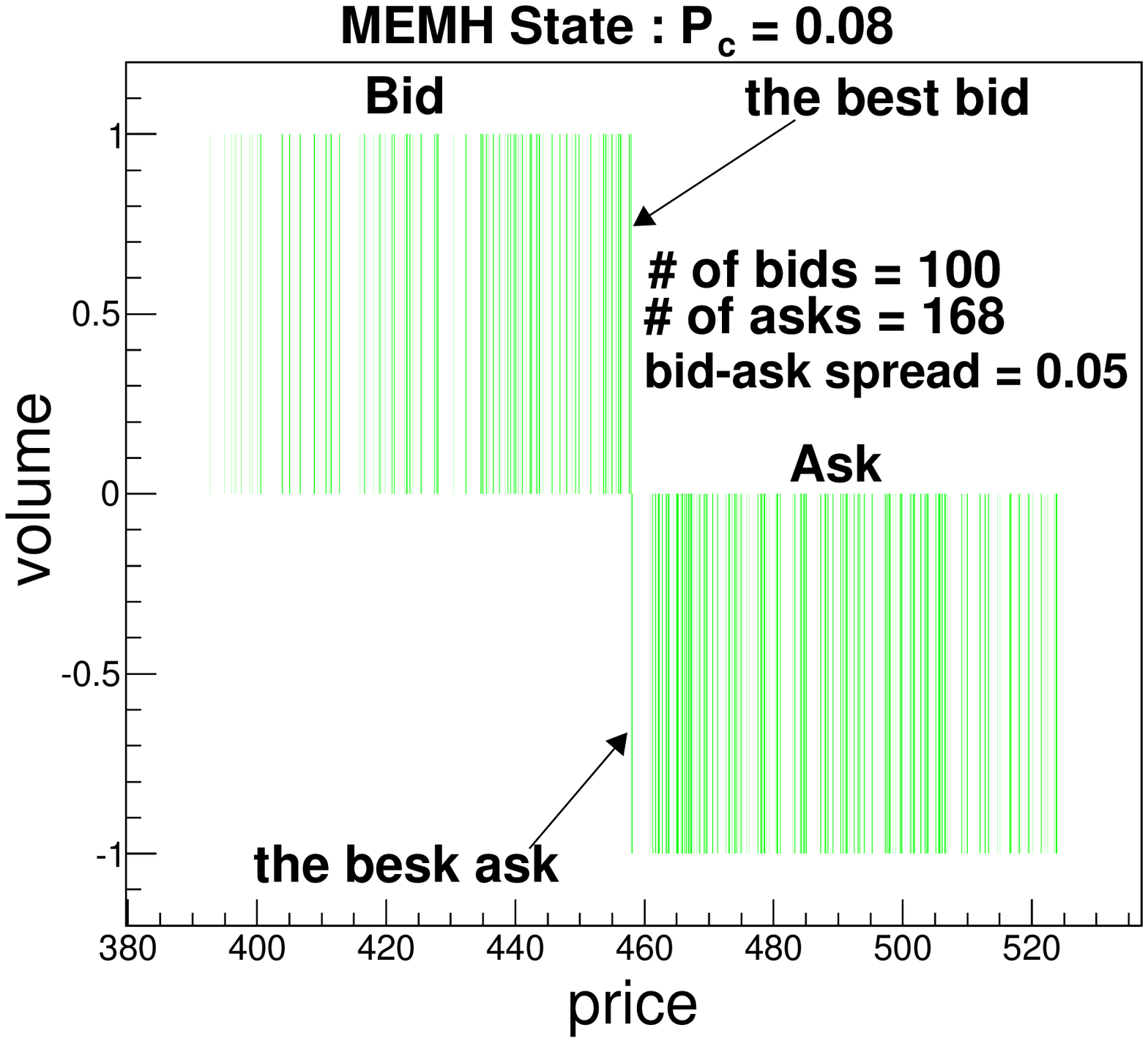}
  }
  \centerline
  {
    \includegraphics[width=.75\textwidth, height=.25\textheight]{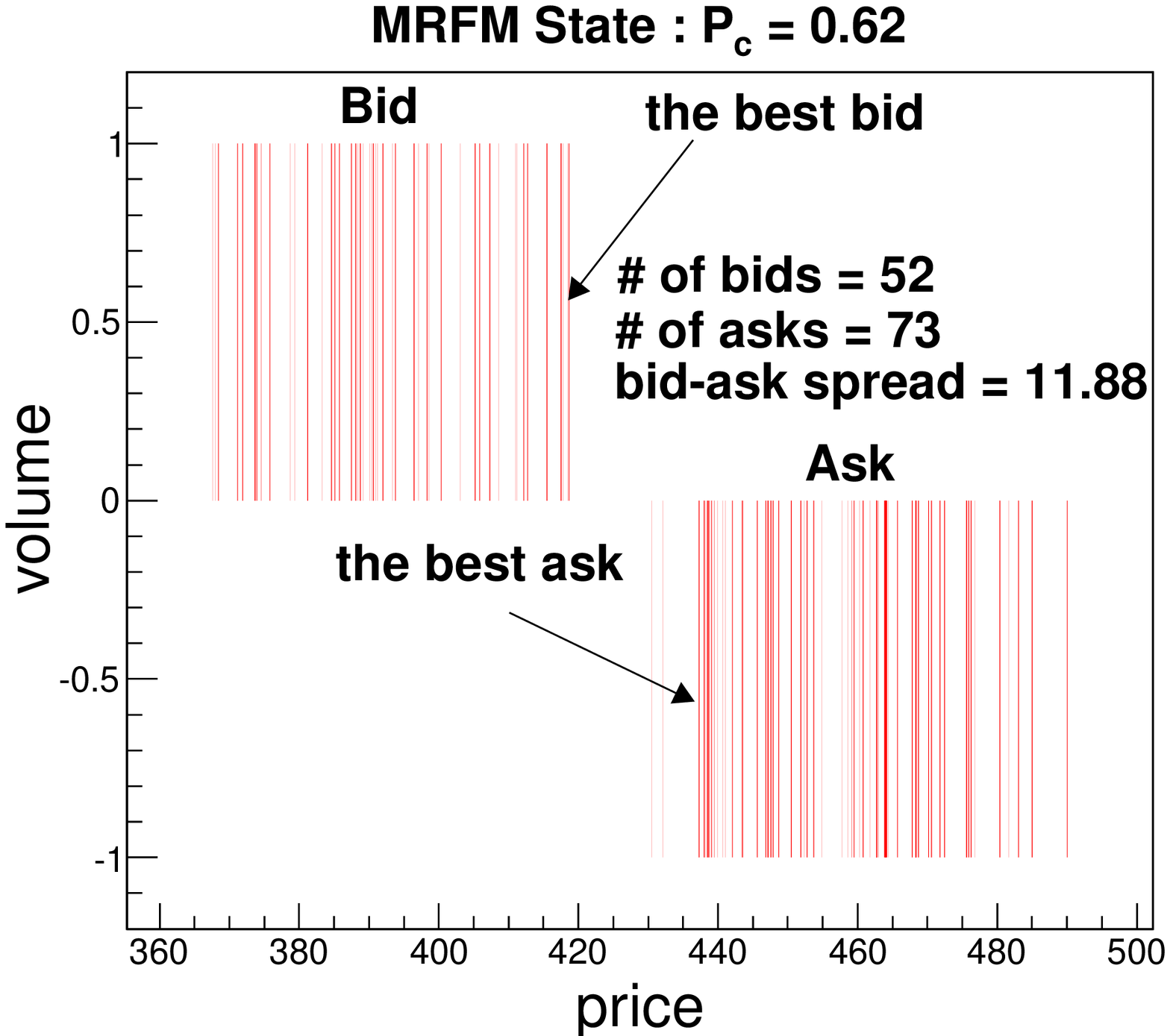}
  }
  \centerline
  {
    \includegraphics[width=.75\textwidth, height=.25\textheight]{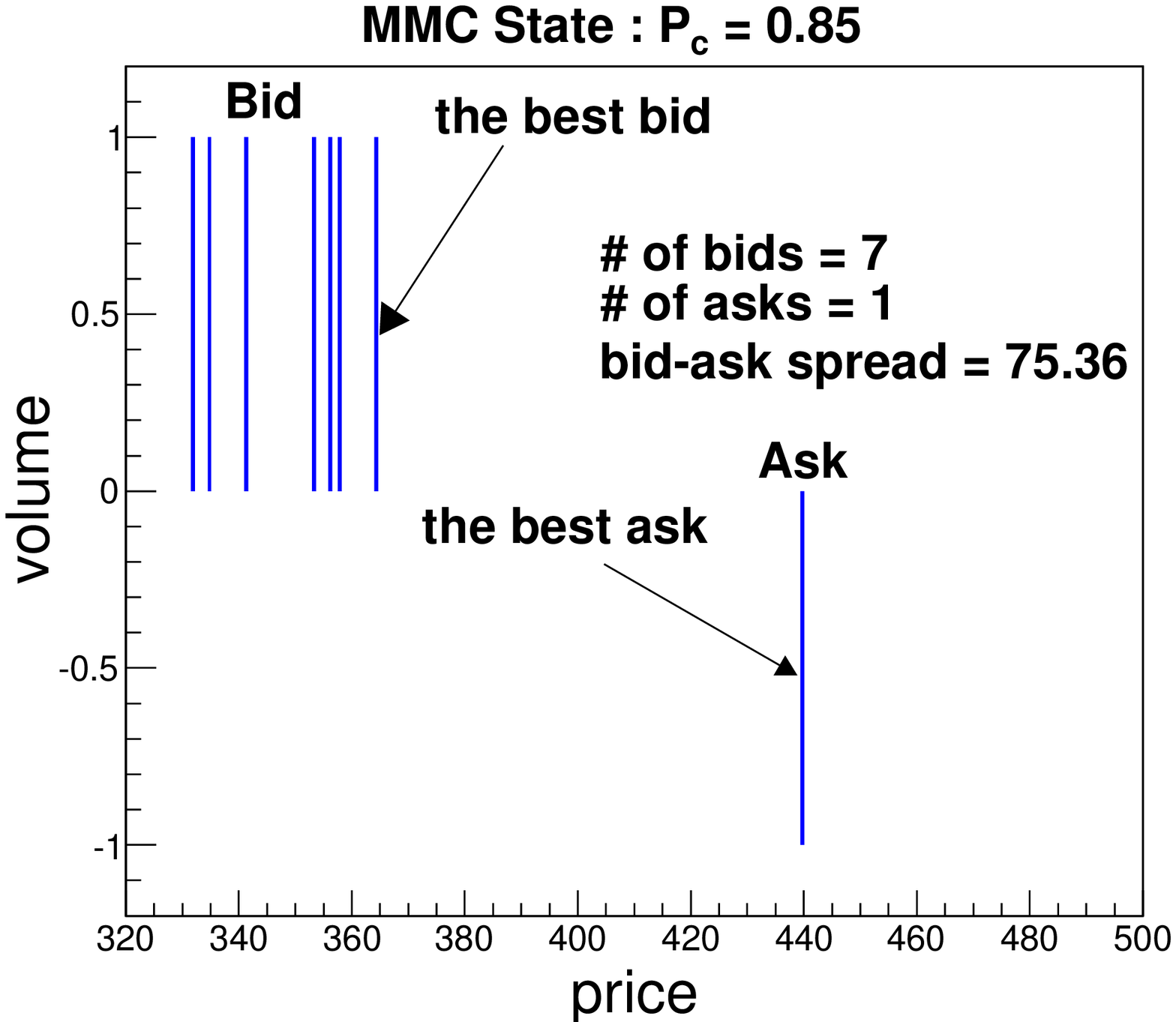}
  }
  \caption{There are snapshots of limit order books in MEMH, MRFM and MMC states. A negative volume indicates the volume of asks. The bid-ask spread is defined by the difference between the best ask, which is the lowest ask in limit orders, and the best bid, which is the highest bid in limit orders.}
  \label{fig:fig4}
\end{figure*}

\begin{figure*}[ht]
  \centerline
  {
    \includegraphics[width=.7\textwidth, height=.4\textheight]{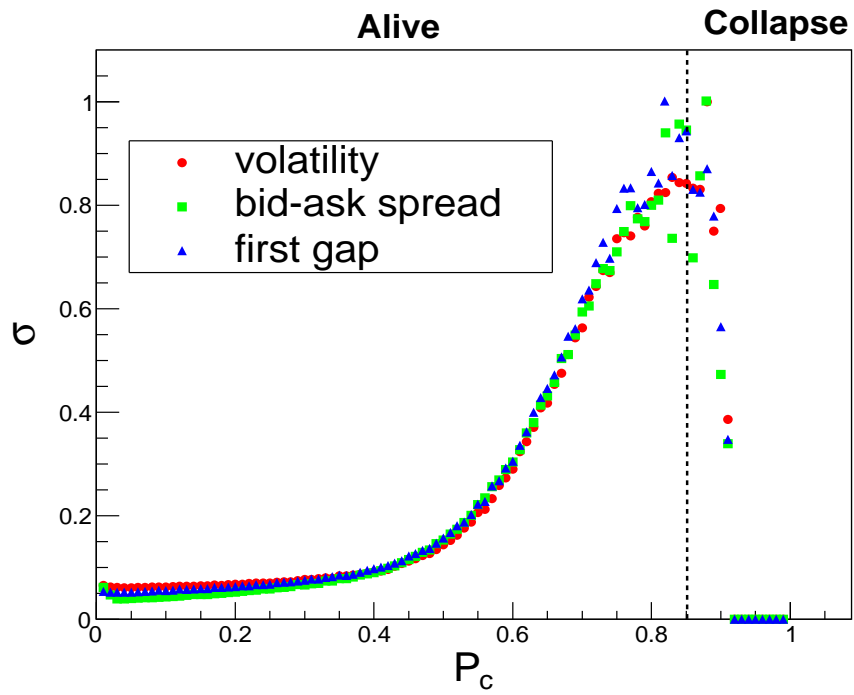}
  }
  \caption{$\sigma$ (Standard deviation) of volatility, the bid-ask spread and the first gap as a function of $P_{c}$. The maximum value of $\sigma$ is normalized by 1.}
  \label{fig:fig5}
\end{figure*}

\begin{figure*}[ht]
  \centerline
  {
    \includegraphics[width=0.5\textwidth, height=0.25\textheight]{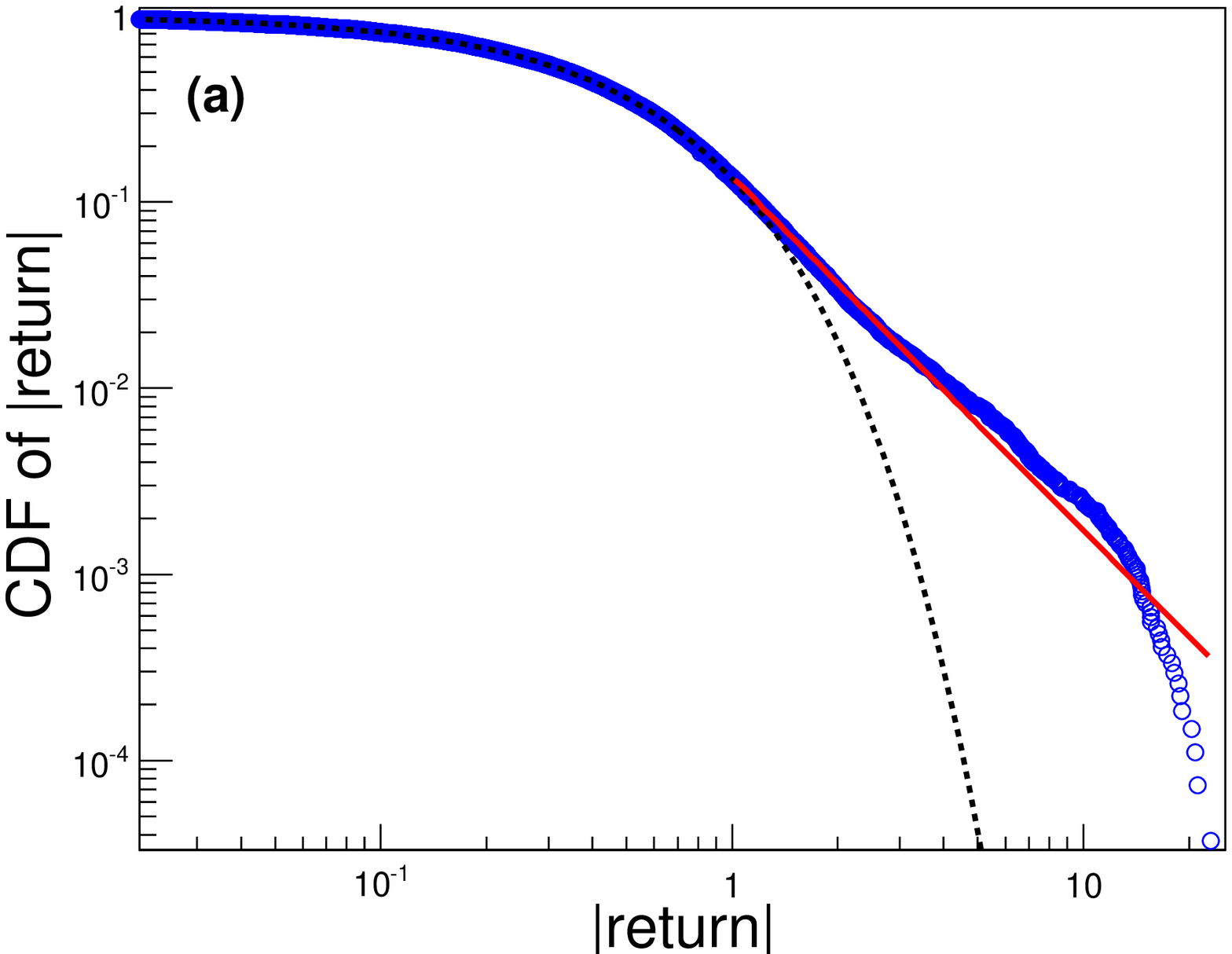}
    \includegraphics[width=0.5\textwidth, height=0.25\textheight]{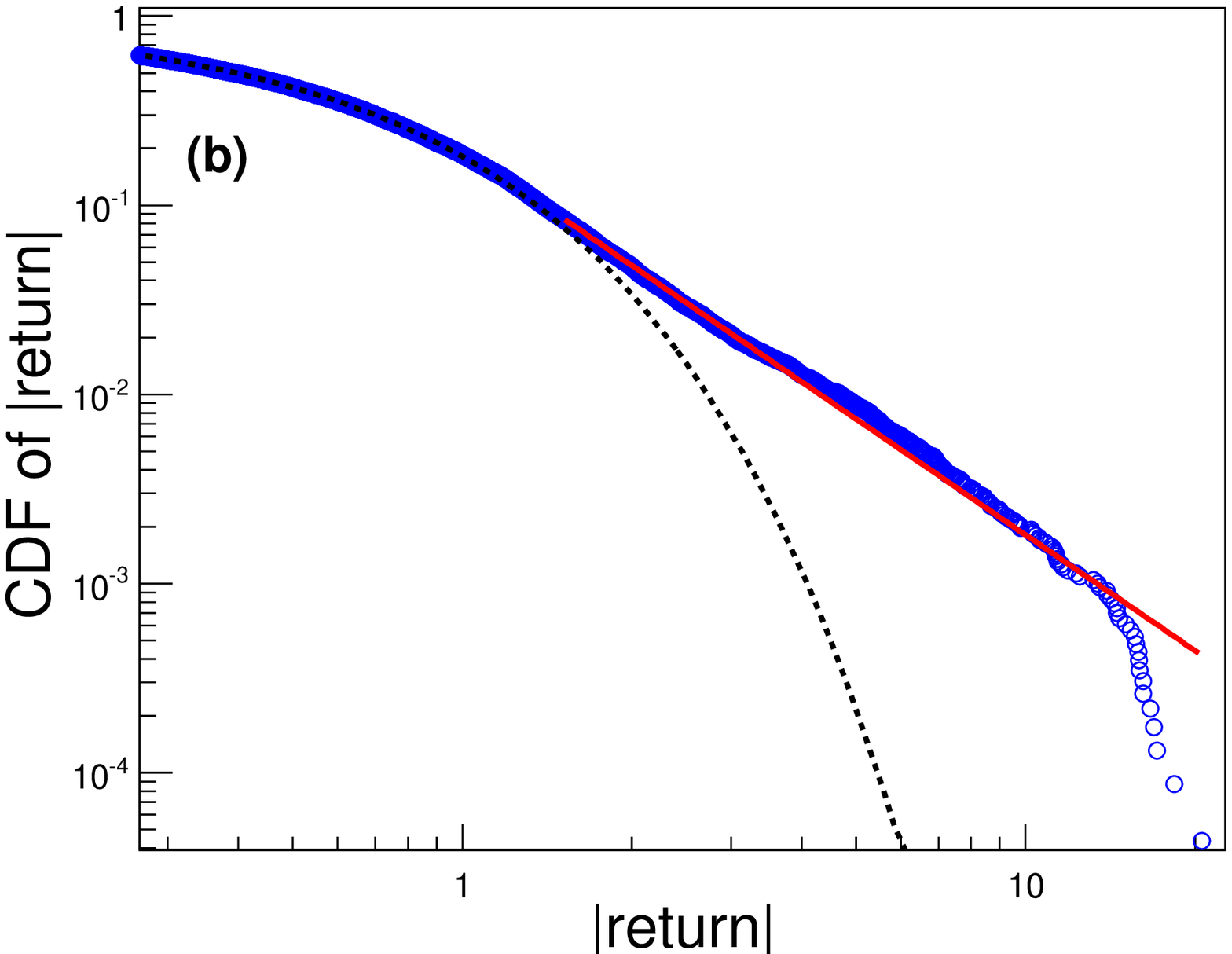}
  }
  \centerline
  {
    \includegraphics[width=0.5\textwidth, height=0.25\textheight]{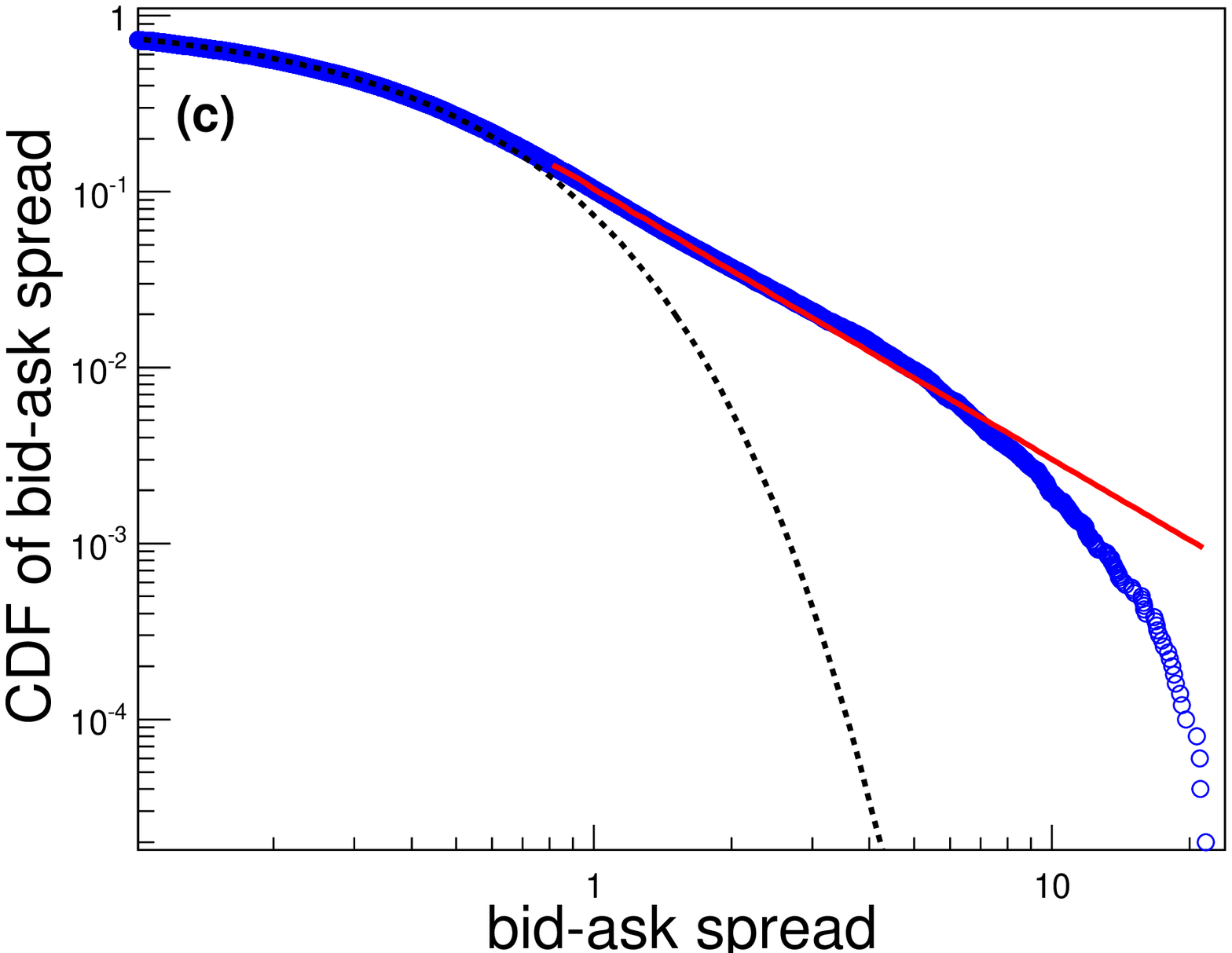}
    \includegraphics[width=0.5\textwidth, height=0.25\textheight]{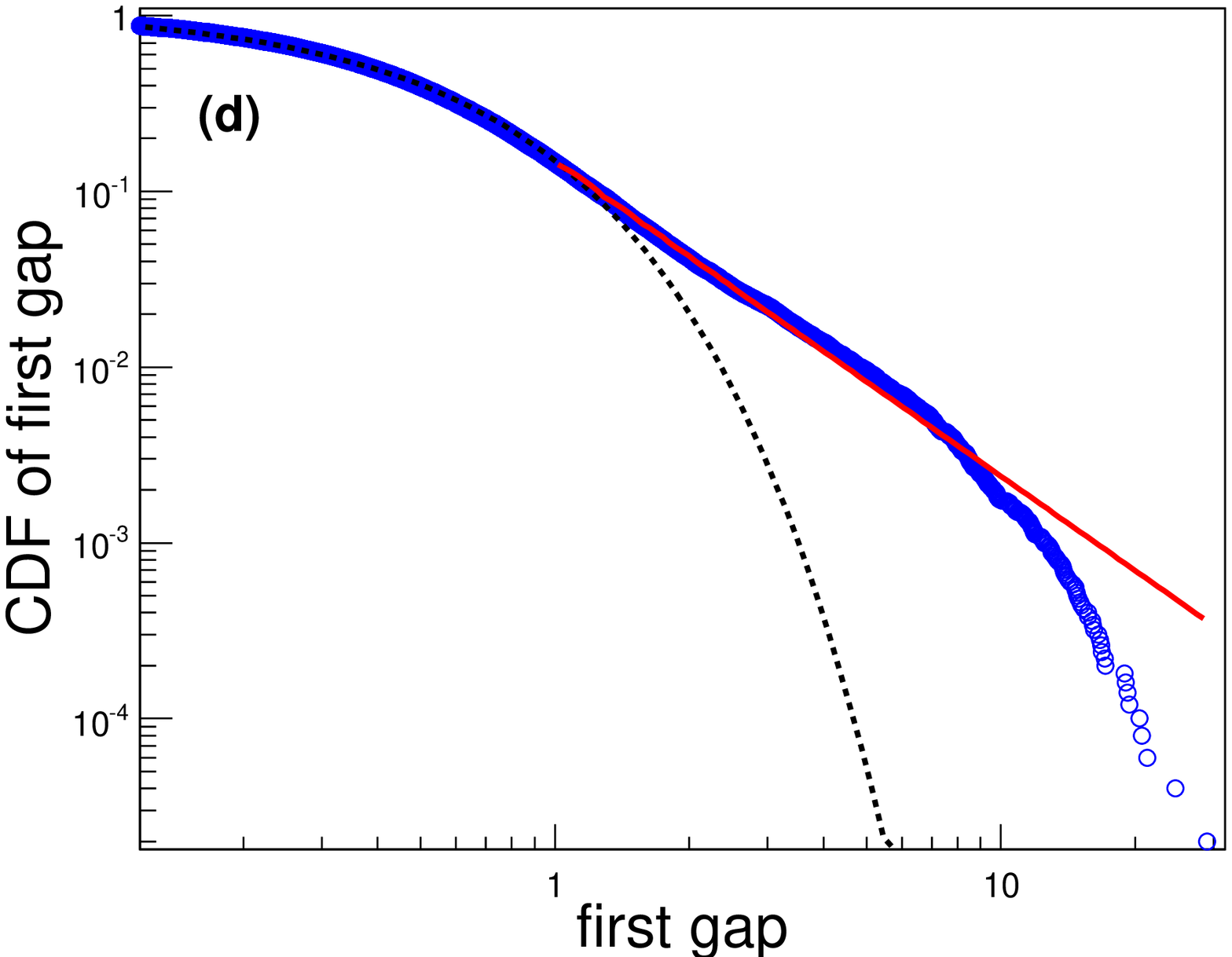}
  }
  \caption{(a)(b)(c)(d)CDFs of positive return, negative return, bid-ask spread and first gap. The dashed black line represents an exponential distribution function. The solid red line represents the fitted power-law function using $y\sim x^{-\alpha}$.The values of the fitted $\alpha$ are summarized in Table \ref{tab:tabA1}}
  \label{fig:figA1}
\end{figure*}

\begin{figure*}[ht]
  \centerline
  {
    \includegraphics[width=0.5\textwidth, height=0.25\textheight]{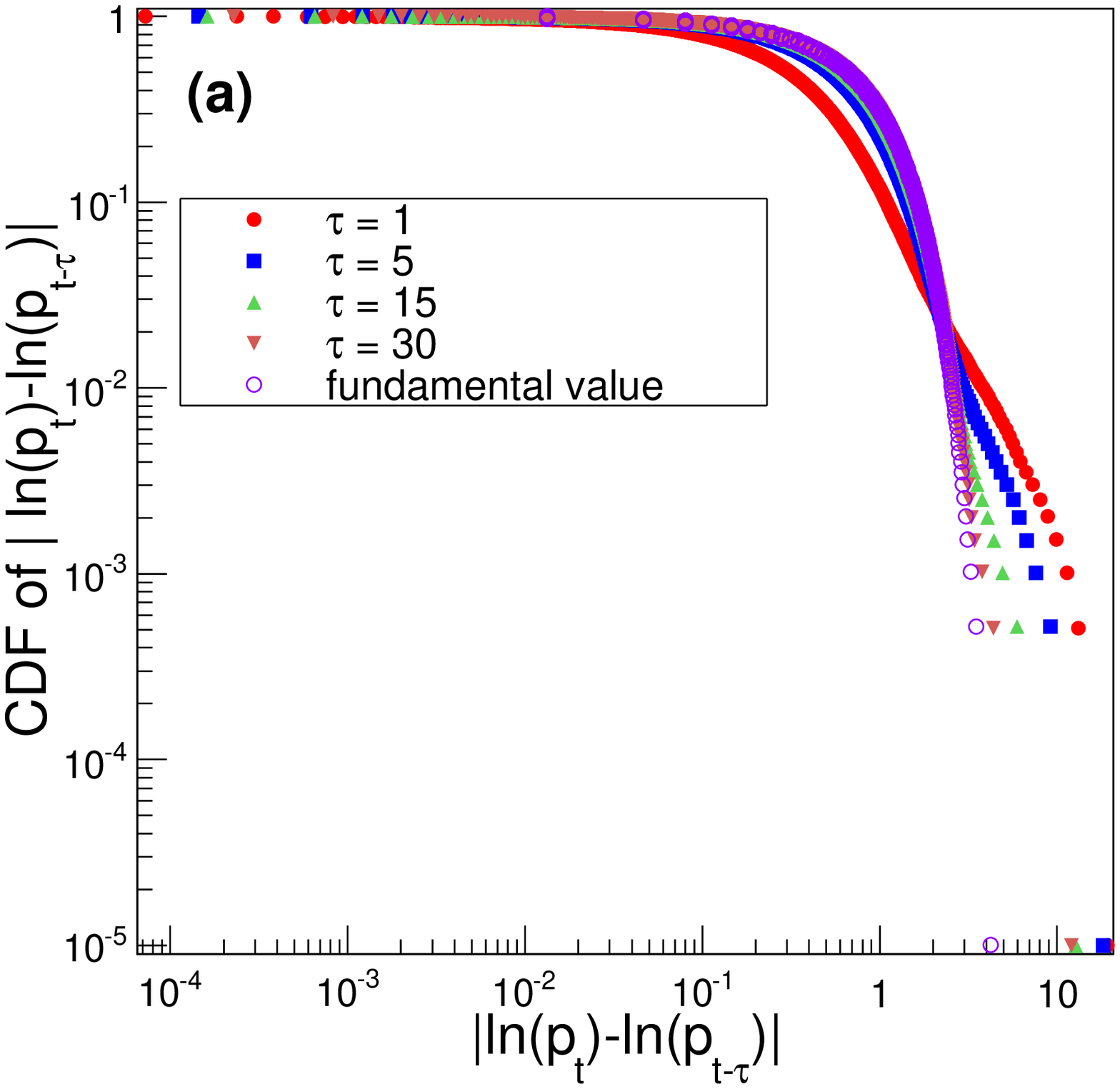}
    \includegraphics[width=0.5\textwidth, height=0.25\textheight]{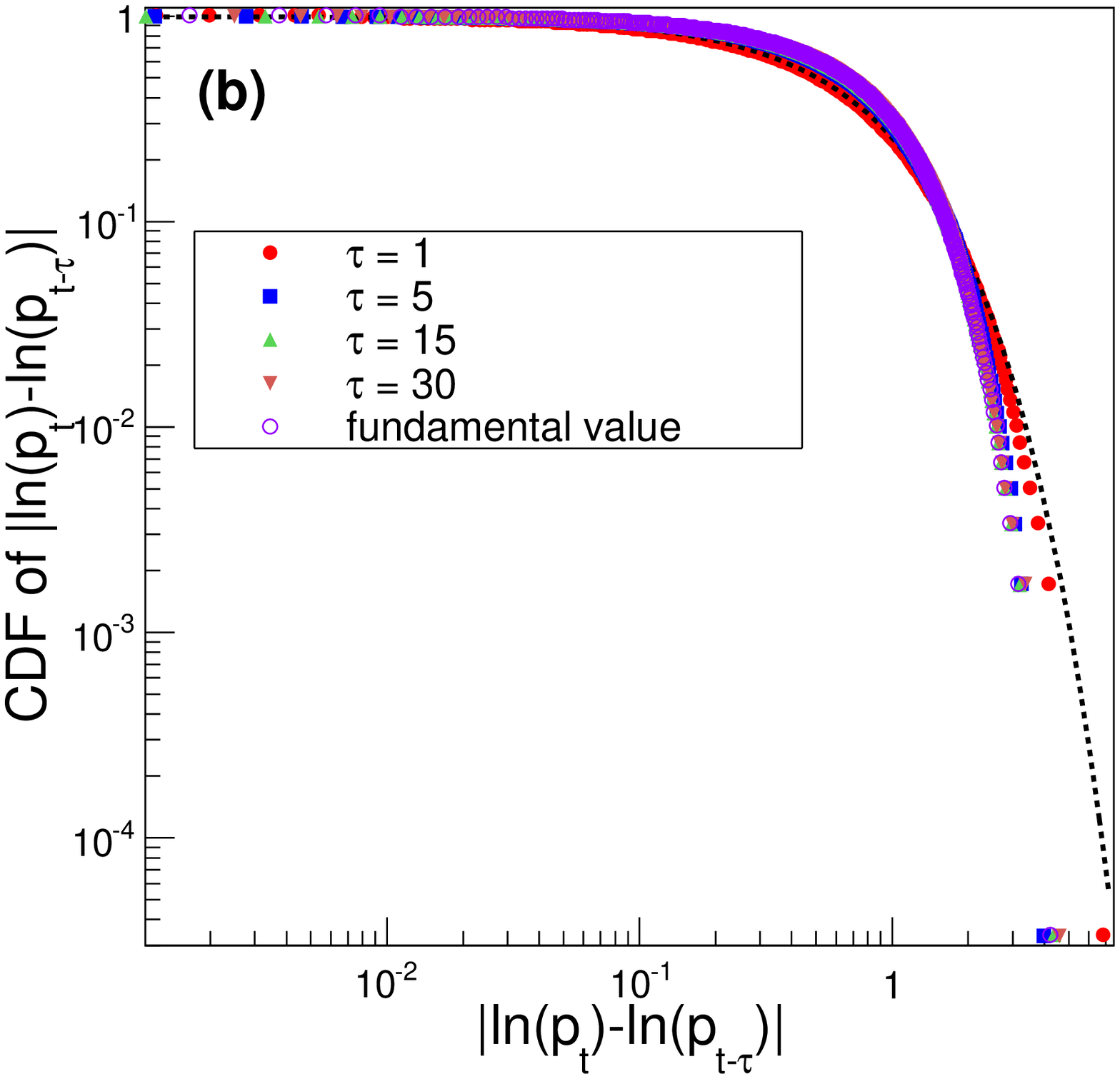}
  }
  \caption{(a) CDF of $|$return$|$. Return is defined by r(t) = $\ln p(t)- \ln p(t-\tau)$. $p(t)$ denotes the market price at time t. (b) the CDF of $|$return$|$  in a homogeneous equilibrium market(i.e., all agents are fundamentalist in the market). The dashed black line represents an exponential distribution function.}
  \label{fig:figA2}
\end{figure*}

\begin{figure*}[ht]
  \centerline
  {
    \includegraphics[width=0.5\textwidth, height=0.25\textheight]{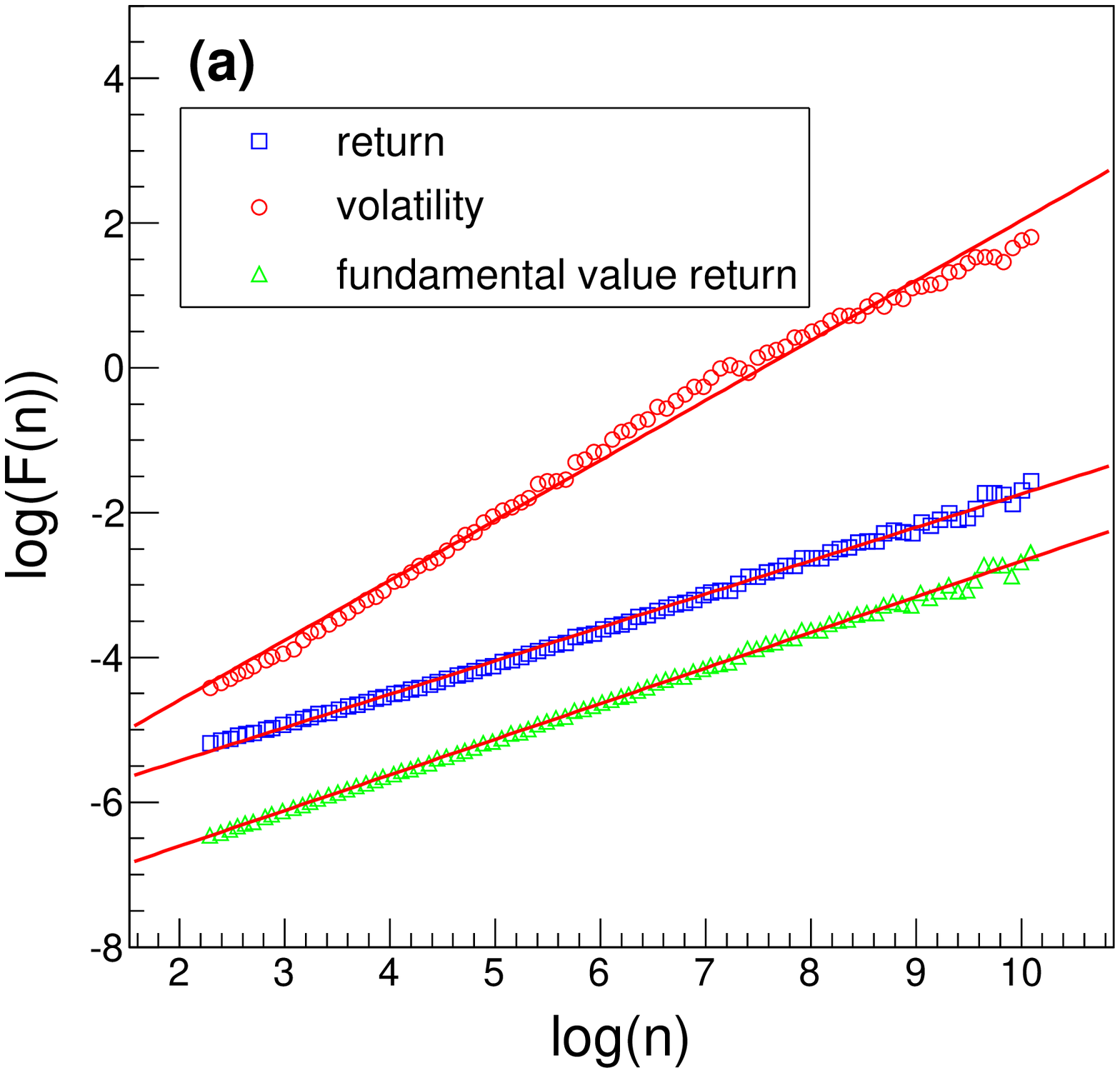}
    \includegraphics[width=0.5\textwidth, height=0.25\textheight]{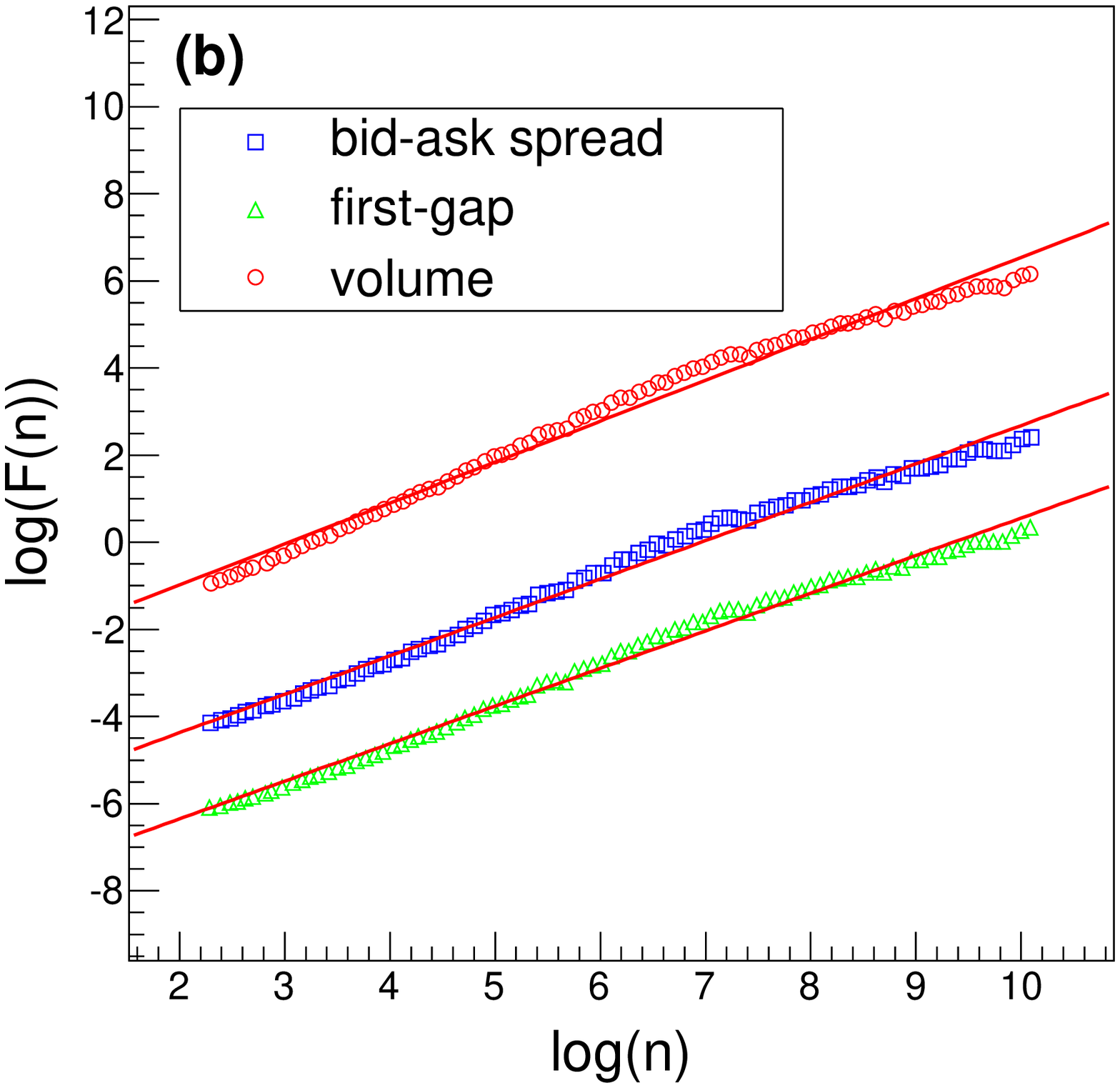}
  }
  \centerline
  {
    \includegraphics[width=0.5\textwidth, height=0.25\textheight]{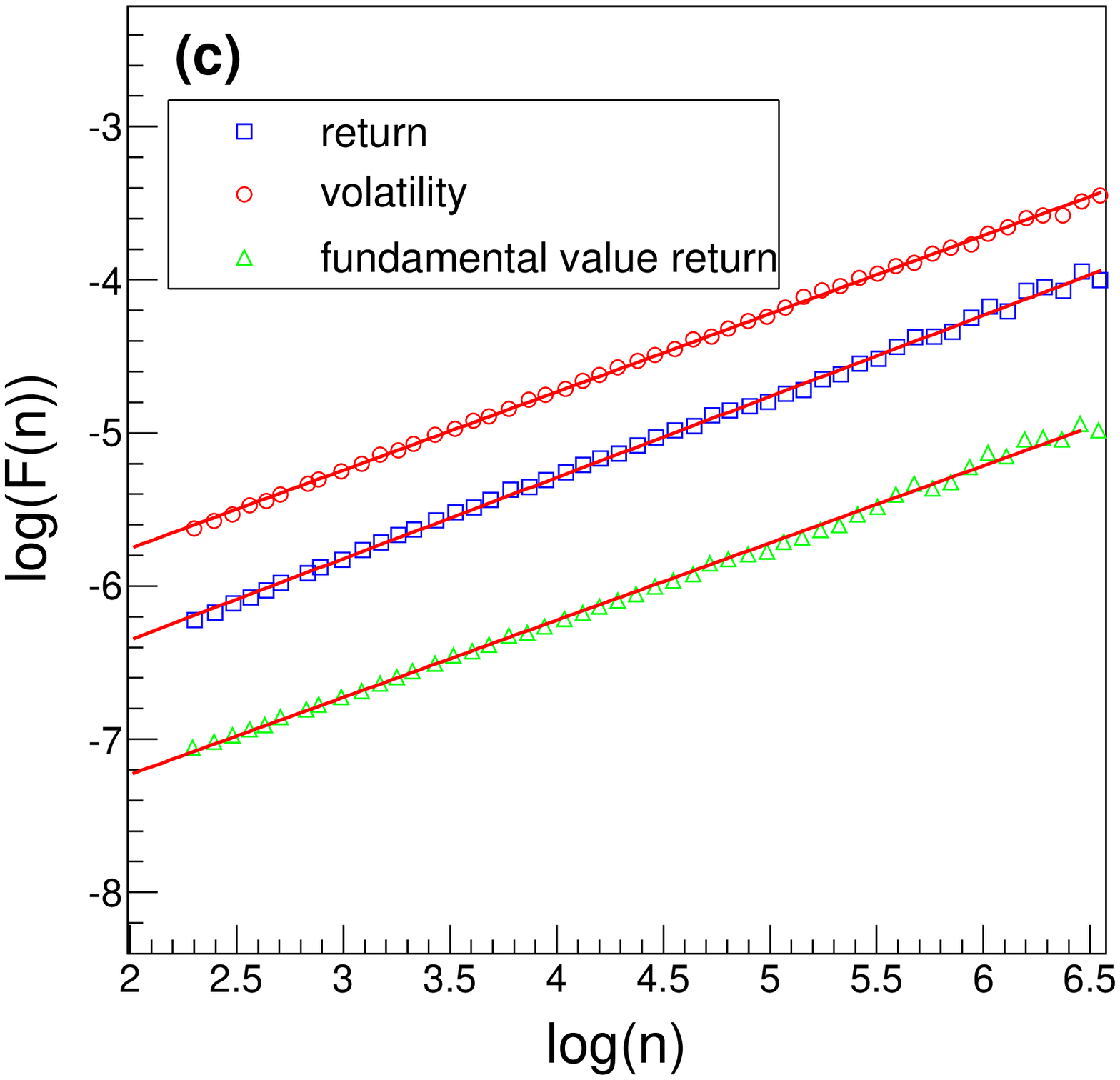}
    \includegraphics[width=0.5\textwidth, height=0.25\textheight]{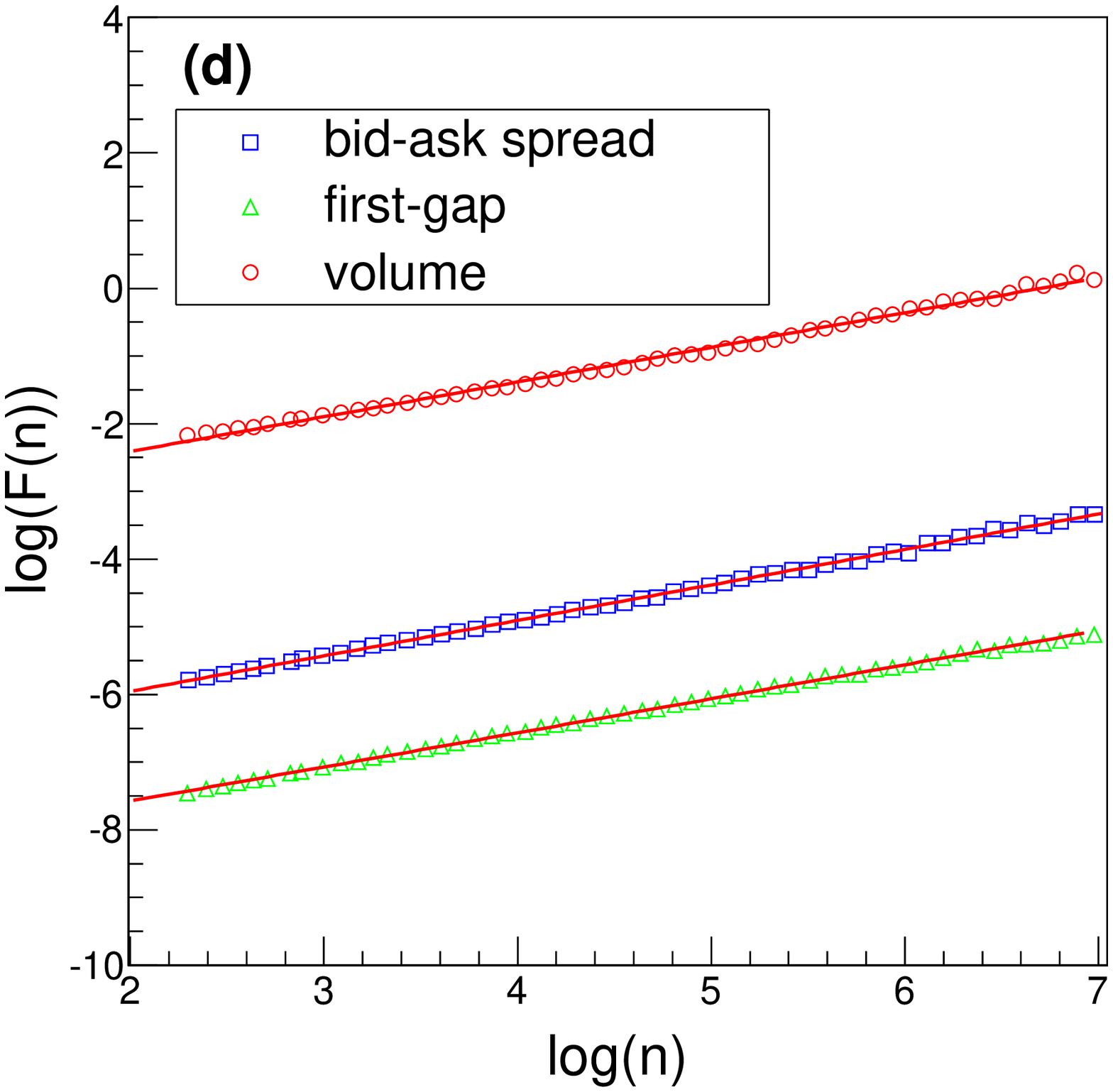}
  }
  \caption{(a)(b) F(n) denotes the rms fluctuation function. The slope of the rms fluctuation function indicates the Hurst exponent. The Hurst exponents measured are summarized in Table \ref{tab:tabA2}.(c)(d). F(n) is the rms fluctuation function in the homogeneous equilibrium market. The Hurst exponents measured in this case are summarized in Table \ref{tab:tabA3}.}
  \label{fig:figA3}
\end{figure*}

\begin{figure*}[ht]
  \centerline
  {
    \includegraphics[width=1.0\textwidth, height=0.7\textheight]{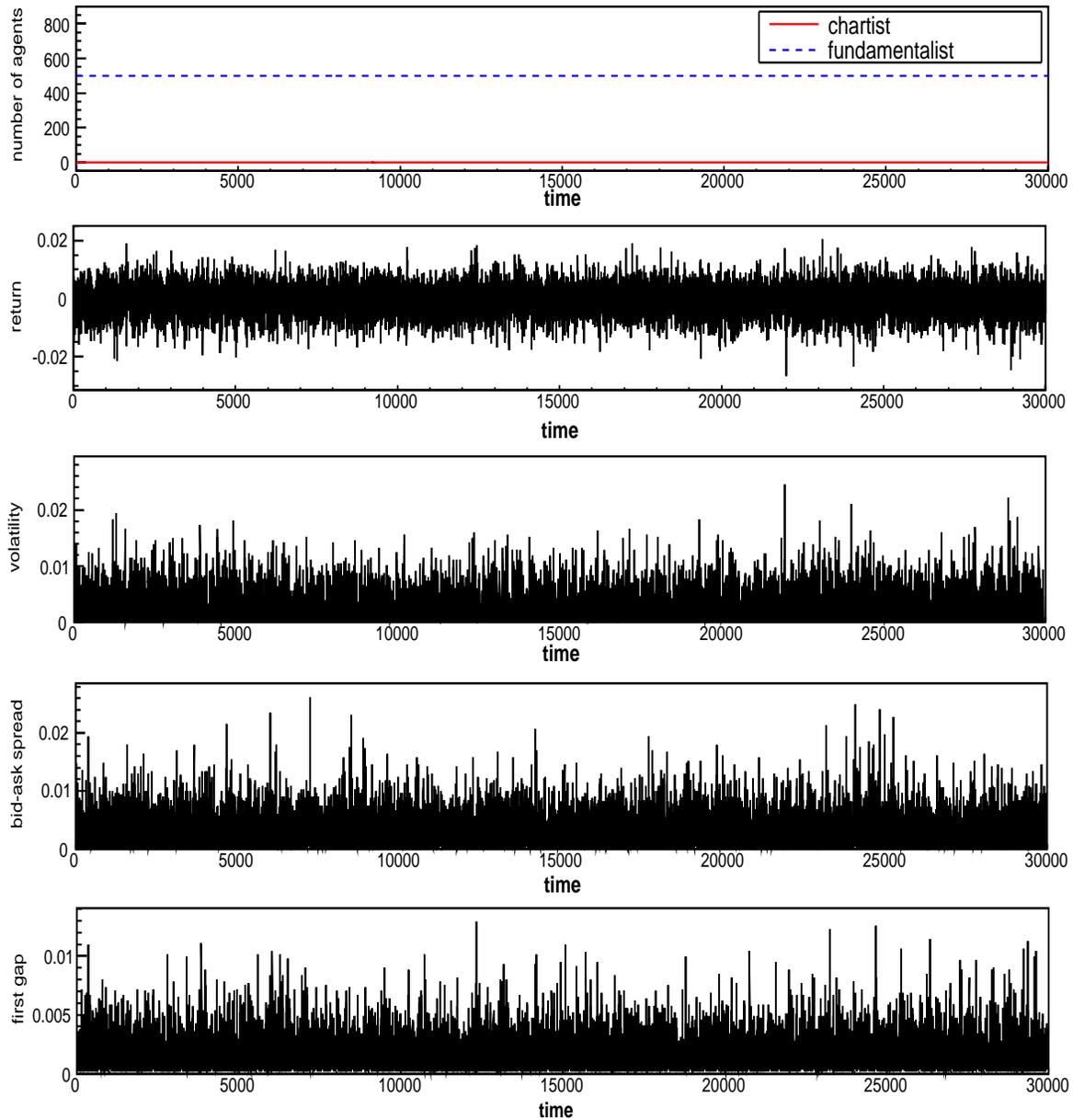}
  }
  \caption{From the top to bottom in figures, the population dynamics of agents' types in the market, return, volatility, bid-ask spread and first gap as a function of time in the homogeneous equilibrium market. In the top figure, the solid red line shows the number of chartists, and the blue dashed line shows the number of fundamentalists in the market.}
  \label{fig:figA4}
\end{figure*}

\begin{figure*}[ht]
  \centerline
  {
    \includegraphics[width=0.5\textwidth, height=0.25\textheight]{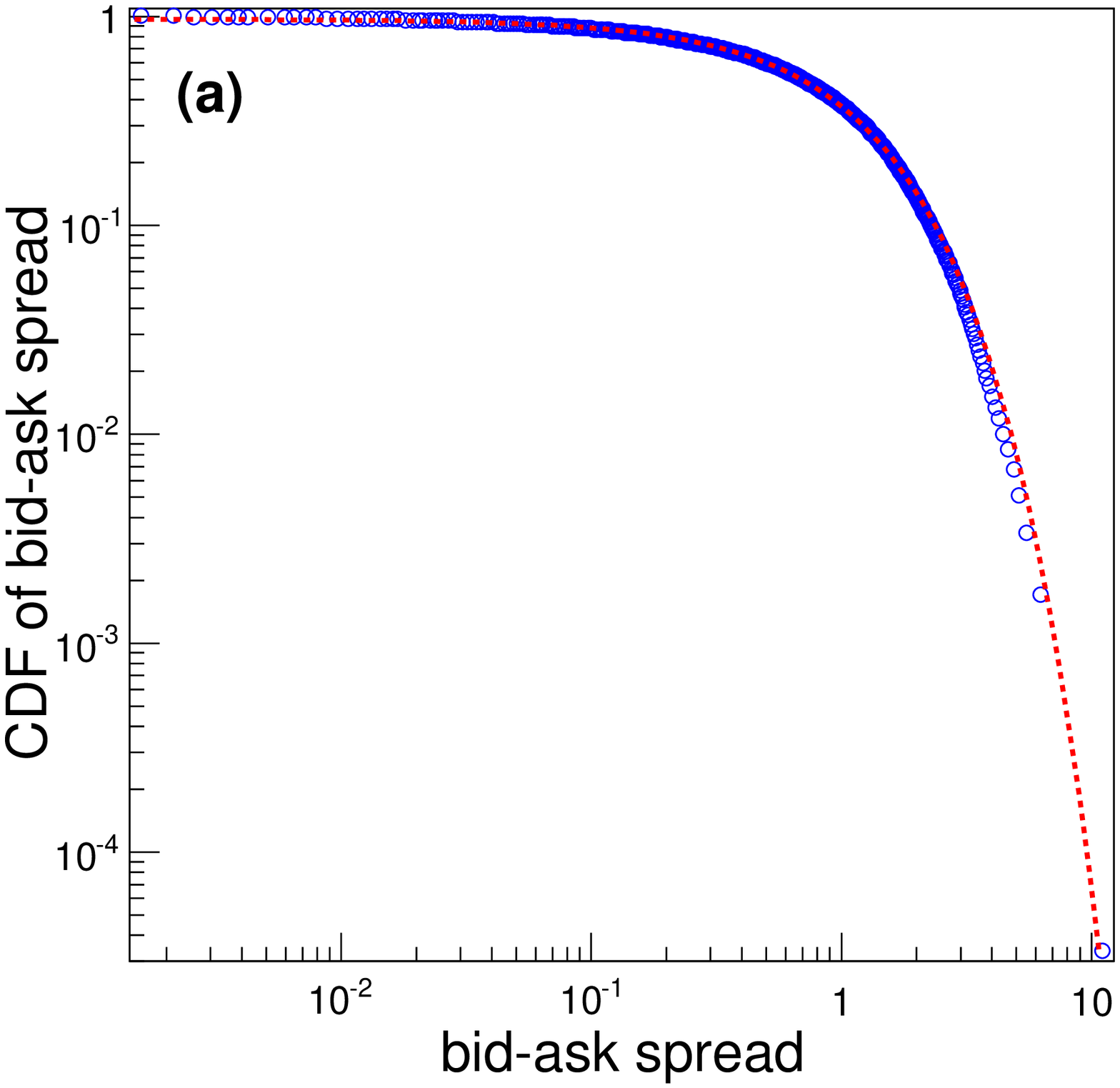}
    \includegraphics[width=0.5\textwidth, height=0.25\textheight]{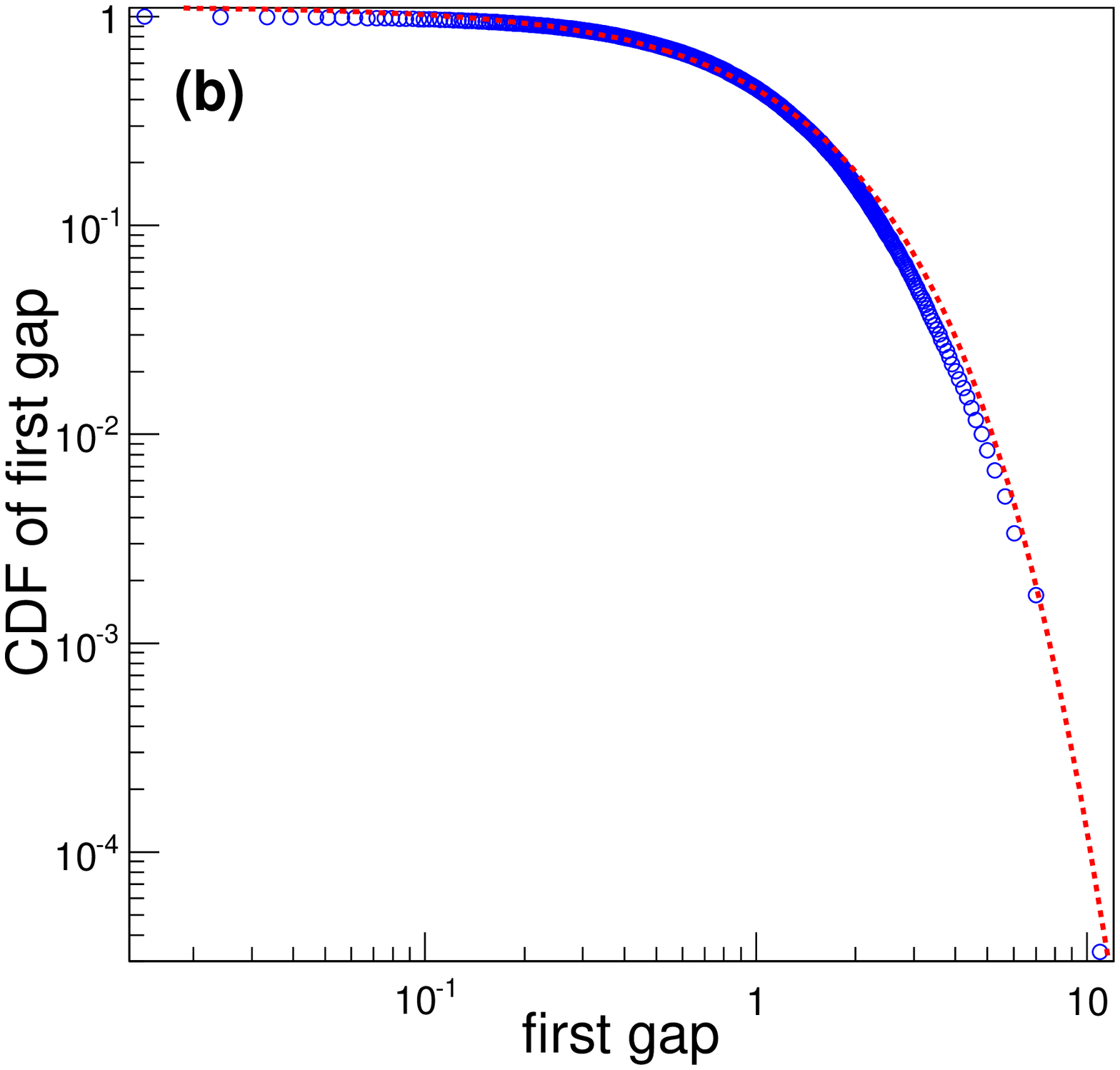}
  }
  \caption{(a)(b) The CDF of the bid-ask spread and the first gap in the homogeneous equilibrium market. The dashed red line shows an exponential distribution function.}
  \label{fig:figA5}
\end{figure*}

\begin{table*}[!ht]
\begin{center}
\begin{tabular}{|c|c|c|}
\hline
 & ADAM & Empirical Result\\
\hline
return(positive tails) &1.89 $\pm$ 0.03  & 2.19 \cite{lillo}\\
\hline
return(negative tails) & 2.03 $\pm$ 0.04  & 2.17 \cite{lillo}\\
\hline
bid-ask spread &1.53 $\pm$ 0.01 & 3.0 \cite{plerou2}\\
\hline
first gap &1.78 $\pm$ 0.02 & 1.5 - 3.0 \cite{farmer}\\
\hline
\end{tabular}
\end{center}
\caption{The fitted results of the power-law function, $y \sim x^{-\alpha}$. $\alpha$ is the power-law exponent. The elements of column ADAM(empirical result) show the results of the ADAM(empirical works).}
\label{tab:tabA1}
\end{table*}

\begin{table*}[!ht]
\begin{center}
\begin{tabular}{|c|c|c|c|}
\hline
 & ADAM & Empirical Result & Memory Type\\
\hline
return &0.46 $\pm$ 0.01 & 0.5 \cite{oh} & No(No)\\
\hline
volatility &0.82 $\pm$ 0.01&0.67 \cite{liu} & Long(Long)\\
\hline
bid-ask spread &0.87 $\pm$ 0.01 & 0.73 $\pm$ 0.01\cite{plerou2} &Long(Long)\\
\hline
first gap &0.87 $\pm$ 0.01& 0.76\cite{lillo} &Long(Long)\\
\hline
FV return & 0.49 $\pm$ 0.01 & & No\\
\hline
volume &0.94 $\pm$ 0.01 & 0.64 - 0.7 \cite{li}& Long(Long)\\
\hline
\end{tabular}
\end{center}
\caption{The Hurst exponent results of the ADAM and the comparison to empirical works. The elements of column ADAM(empirical result) show the results of the ADAM(empirical works). The words in parenthesis at Memory Type indicate the memory type observed in empirical works.[FV : fundamental value]}
\label{tab:tabA2}
\end{table*}

\begin{table*}[!ht]
\begin{center}
\begin{tabular}{|c|c|c|}
\hline
 & ADAM & Memory Type\\
\hline
return &0.52 $\pm$ 0.01 & No\\
\hline
volatility &0.51 $\pm$ 0.01& No\\
\hline
bid-ask spread &0.52 $\pm$ 0.01 & No\\
\hline
first gap &0.50 $\pm$ 0.01 & No\\
\hline
FV return & 0.50 $\pm$ 0.01 & No\\
\hline
volume &0.51 $\pm$ 0.01 & No\\
\hline
\end{tabular}
\end{center}
\caption{Hurst exponents in the homogeneous equilibrium market of the ADAM. There is no memory in market microstructures in this case.[FV : fundamental value]}
\label{tab:tabA3}
\end{table*}

\end{document}